\documentclass{ws-procs11x85}

\usepackage{pdfpages}

\begin{document}

% RBC: NEW TITLE to incorporate ADR (PSB-Reviewer 3)
%LMR: shortened it a little
%\title{Monitoring Potential Drug Interactions via Network Analysis of\\ Instagram User Timelines}
%\title{Monitoring Potential Drug Interactions and Reactions via Network Analysis of Instagram User Timelines}
\title{Monitoring Potential Drug Interactions and Reactions via Network Analysis of Instagram User Timelines}

\author{Rion Brattig Correia$^{1,2}$, Lang Li$^3$ and Luis M. Rocha$^{1,4,*}$}

\address{
$^1$School of Informatics \& Computing, Indiana University,\\
Bloomington, IN 47408 USA\\
$^*$rocha@indiana.edu}

\address{$^2$CAPES Foundation, Ministry of Education of Brazil,\\
Bras\'{i}lia, DF 70040-020, Brazil}

\address{$^3$Department of Medical and Molecular Genetics, Indiana University School of Medicine,\\
Indianapolis, IN 46202 USA}

\address{$^4$Instituto Gulbenkian de Ci\^{e}ncia,\\
Oeiras 2780-156, Portugal}

%\address{
%$^1$School of Informatics \& Computing, Indiana University,\\
%Bloomington, IN 47408 USA\\
%$^*$rocha@indiana.edu\\
%$^2$CAPES Foundation, Ministry of Education of Brazil,\\
%Bras\'{i}lia, DF 70040-020, Brazil\\
%$^3$Department of Medical and Molecular Genetics, Indiana University School of Medicine,\\
%Indianapolis, IN 46202 USA\\
%$^4$Instituto Gulbenkian de Ci\^{e}ncia,\\
%Oeiras 2780-156, Portugal}

\begin{abstract}
Much recent research aims to identify evidence for Drug-Drug Interactions (DDI) and Adverse Drug reactions (ADR) from the biomedical scientific literature.
In addition to this ``Bibliome'', the universe of social media provides a very promising source of large-scale data that can help identify DDI and ADR in ways that have not been hitherto possible.
Given the large number of users, analysis of social media data may be useful to identify under-reported, population-level pathology associated with DDI, thus further contributing to improvements in population health. Moreover, tapping into this data allows us to infer drug interactions with natural products---including cannabis---which constitute an array of DDI very poorly explored by biomedical research thus far.

Our goal is to determine the potential of \textit{Instagram} for public health monitoring and surveillance for DDI, ADR, and behavioral pathology at large. Most social media analysis focuses on \emph{Twitter} and \emph{Facebook}, but \emph{Instagram} is an increasingly important platform, especially among teens, with unrestricted access of public posts, high availability of posts with geolocation coordinates, and images to supplement textual analysis.

%The current Instagram penetration rate among internet users in the United States is close to 20 percent with this number projected to surpass 50 million in 2016. It also ranks second among teens (12-24) in social media usage.

Using drug, symptom, and natural product dictionaries for identification of the various types of DDI and ADR evidence, we have collected close to 7000 user timelines spanning from October 2010 to June 2015. We report on 1) the development of a monitoring tool to easily observe user-level timelines associated with drug and symptom terms of interest, and 2) population-level behavior via the analysis of co-occurrence networks computed from user timelines at three different scales: monthly, weekly, and daily occurrences. Analysis of these networks further reveals 3) drug and symptom direct and indirect
associations with greater support in user timelines,
as well as 4) clusters of symptoms and drugs revealed by the collective behavior of the observed population.
%, and 4) substantial redundancy in networks and timelines, allowing us to considerably reduce the amount of potential interaction and user timeline data delivered to human and automated analysis.

This demonstrates that \emph{Instagram} contains much drug- and pathology specific data for public health monitoring of DDI and ADR, and that complex network analysis provides an important toolbox to extract health-related associations and their support from large-scale social media data.

\end{abstract}


\bodymatter

%
% Introduction
%
\section{Introduction}
\label{sec:introduction}

The analysis of social media data has recently allowed unprecedented access to collective human behavior. The new field of Computational Social Science has brought together Informatics and Complex Systems methods to study society via social media and online data in a quantitative manner not previously possible.
From studying social protest \cite{Varol:2014} to predicting the Stock Market \cite{Bollen:2011}, most of the work has focused on \emph{Twitter}---though \emph{Facebook}\cite{Bakshy:2015}
%, \emph{Flickr}\cite{Cha:2012}
and \emph{Instagram} \cite{Ferrara:2014} have also received some attention lately.
This approach shows great promise in monitoring public health, given the ability to measure the behavior of a very large number of human subjects\cite{Kautz:IJCAI:2013}.
For instance,
%\emph{Google} searches have been shown to correlate with dengue spread in tropical zones \cite{Chan:2011} and
several studies have shown that social media analysis is useful to track and predict influenza spread \cite{Kautz:IJCAI:2013,Signorini:2011,Sadilek:2012}, as well as the measurement of depression \cite{Choudhury:2013}.
In particular, the potential for adverse drug reaction (ADR) extraction from \emph{Twitter} has been recently demonstrated \cite{ahmed2015,gonzalez2015}.

There is still, however, much work to be done in order to fulfill the potential of social media in the monitoring of public health.
For instance, analysis of social media data may be useful to identify under-reported pathology, particularly in the case of conditions associated with a perceived social stigma, such as mental disorders \cite{pescosolido2015stigma}.
Given access to an extremely large population, it is reasonable to expect that social media data may provide early warnings about potential drug-drug interactions (DDI) and ADR \cite{ahmed2015}.
These unprecedented windows into collective human behavior may also be useful to study the use and potential interactions and effects of natural products---including cannabis.
The pharmacology of such products constitute an array of DDI and ADR very poorly explored by biomedical research so far, and thus an arena where social media mining could provide important novel discoveries and insight.

% RBC (PSB Reviewer 1) = changed "All work" for "Most work"
Most work on social media pertaining to public health monitoring that we are aware of has relied on data from \emph{Twitter} or \emph{Facebook}. However, \emph{Instagram} is an increasingly important platform, with unrestricted access of public posts, high availability of posts with geolocation coordinates, and images to supplement textual analysis.
While Instagram has been used to qualitatively observe the type of content people post regarding health situations such as Ebola outbreaks \cite{Seltzer:2015}, its potential for large-scale quantitative analysis in public health has not been established.
%
%The current \emph{Instagram} penetration rate among internet users in the United States is close to 20 percent with this number projected to surpass 50 million in 2016. It also ranks second among teens (12-24) in social media usage.
%
\emph{Instagram} currently has more than 300 million users \cite{InstagramBlog:2014}. It surpasses \emph{Twitter} and \emph{Facebook} for preferred social network among teens (12-24) in the US. In 2014 there were approximately more than 64 million active users in the US and this number is to surpass 111 million in 2019 \cite{Statista:2015}.
% RBC (PSB Reviewer 1) = Added phrase
%Still the platform is very under-researched with only a few works focusing on public health \cite{Seltzer:2015}.
%
%LMR: I moved and edited this above.
%
Therefore, our goal here is to explore the potential of this very important social media platform for public health monitoring and surveillance of DDI, ADR, and behavioral pathology at large.
% RBC (PSB Reviewer 1) = Added phrase
% LMR: Edited it a bit.
Specifically, we use literature mining and network science methods to automatically characterize and extract temporal signals for DDI and ADR from a sub-population of Instagram users.

We focused on posts and users with mentions of drugs known to treat depression (e.g. \texttt{fluoxetine}).
The methodology developed can be easily replicated for different clinical interests (e.g. epilepsy drugs). The goal is to show that Instagram is a very rich source of data to study drug interactions and reactions that may arise in a clinical context of choice, and not depression per se.
Using four different multi-word dictionaries (drug and pharmacology, natural products, cannabis, and ADR terminology), we have collected close to 7000 user timelines spanning from October 2010 to June 2015.
%
%RBC: Phrase changed. (PSB-REVIEWER 1)
%We have analyzed mentions and co-mentions in three temporal different scales---monthly, weekly and daily.
We analyzed co-mentions in three distinct time-windows: monthly, weekly and daily. This allows the potential extraction of ADR and DDI that manifest at different time scales.
From this data, we demonstrate that \emph{Instagram} user timelines contain substantial data of interest to characterize DDI, ADR, and natural product use.
To explore this data we have developed a monitoring tool to easily observe user-level timelines associated with drug and symptom terms of interest, which we describe below.
To explore population-level associations at the different temporal scales, we compute knowledge networks that our previous work has shown to be useful for automated fact-checking\cite{Ciampaglia:2015}, protein-protein interaction extraction %\cite{Verspoor:2005,Abi-Haidar:2008}
\cite{Abi-Haidar:2008}, and recommender systems \cite{Rocha:MyLibraryLANL:2005,Simas:2015}.
%RBC: Phrase added. (PSB-REVIEWER 1)
To illustrate the potential of data-driven, population-level associations, we use spectral methods to reveal network modules of symptoms and drugs, for instance those involved in psoriasis pathology.
Our \emph{Instagram} analysis relies on
%a very different and novel study of redundancy in the topology of complex networks\cite{Simas:2015}.
the distance closure of complex networks \cite{Simas:2015} built at distinct time resolutions, which is a novel development from related approaches to uncover ADR in \emph{Twitter} \cite{ahmed2015}.
%
%This approach allows us to uncover the essential drug and symptom associations and user timelines that preserve shortest path computations in the knowledge networks. This allows us to obtain important direct and indirect (latent) associations in the data, as well as remove many redundant associations from the data.

%
% Method
%
\section{Data and Methods}
\label{Sec:Data_Methods}

We harvested from \emph{Instagram} all posts containing hashtags that matched 7 drugs known to be used in the treatment of depression (\# posts): \texttt{fluoxetine} (8,143), \texttt{sertraline} (574),  \texttt{paroxetine} (470), \texttt{citalopram} (426), \texttt{trazodone} (227), \texttt{escitalopram} (117), and \texttt{fluvoxamine} (22).
Synonyms were resolved to the same drug name according to \emph{DrugBank} \cite{DrugBank}; for instance, \texttt{Prozac} is resolved to \texttt{fluoxetine}, see supporting information (SI) for table of  synonyms used.
%
%LMR: in the future we should use also and commercial brand names as in the precsription products field of DrugBank.
%
This resulted in a total of 9,975 posts from 6,927 users, whose complete timelines, spanning the period from October 2010 to June 2015, were collected.
%in a document-oriented database for fast access.
%
In total, these timelines contain $5,329,720$ posts, which is the depression timeline dataset we analyze below.

A subset of a previously developed pharmacokinetics ontology \cite{Wu2013} was used to obtain a drug dictionary.
The full ontology contains more than 100k drugs, proteins and pharmacokinetic terms. Here we used only names of FDA-approved drugs, along with their generic name and synonyms, resulting in 17,335 drug terms.
The natural product (NP) dictionary was built using terms from the list of herbal medicines and their synonyms provided by MedlinePlus\cite{HerbsMedline}. It contains 179 terms (see SI).
The Cannabis dictionary was assembled by searching the web for terms known to be used as synonyms for cannabis, resulting in 26 terms (see SI) optimized for precision and recall on a subset of posts  (data not shown).
The symptom dictionary was extracted from BICEPP \cite{Lin:2011} by collecting all entities defined as an Adverse Effect, with a few manual edits to include more synonyms; it is comprised of 250 terms.

%
%Eventually we do need to improve our dictionaries a lot.

Timeline posts were tagged with all dictionary terms (n-grams)
for a total of 299,312 matches.
Uppercase characters were converted to lowercase, and hashtag terms were treated like all other harvested text for the purpose of dictionary matches.
We found matches for 414 drugs, 133 of which with more than 10 matches. These numbers are 148/99 and 74/46 for symptoms and NP, respectively, for a total of 636 terms.
%
%LMR: We should probably be more conservative and use only the terms that occurred x times to build the networks. I would say x=5 or 10...
%
This is a substantial number of dictionary terms, given that only 7 drugs prescribed for depression were used to harvest the set of timelines.
The top 25 matches for each dictionary are provided in SI.
Notice that the term `\texttt{depression}' was removed because of its expected high appearance. Matches in the cannabis dictionary (e.g. 420, marijuana, hashish) were aggregated into the term ‘cannabis’ to be treated as a NP.
%
% RBC: To save space I've transformed this table into a paragraph
The top 10 mentions are (counts shown): \texttt{cannabis} (66,540),
%
%\texttt{depression} (25,020),
%
\texttt{anorexia} (26,872), \texttt{anxiety} (26,309), \texttt{pain} (15,677), \texttt{suicide} (11,616), \texttt{mood} (11,532), \texttt{fluoxetine} (9,961), \texttt{suicidal} (8,909), \texttt{ginger} (7,289), \texttt{insomnia} (5,917).

Given the set $X$  of all matched terms ($|X|=636$),
we first compute a symmetric co-occurrence graph $R_w(X)$ for time-window resolutions $w =$ 1 month, 1 week and 1 day.
These graphs are easily represented by adjacency matrices $R_w$, where entries $r_{ij}$ denote the number of time-windows where terms $x_i$ and $x_j$ co-occur, in all user timelines.
%
%$r_{ii}$ denotes the number of time-windows the term $x_i$ occurs in.
%
A matrix $R_w$ is computed for each time-window resolution independently.
To obtain a normalized strength of association among the set of terms $X$, we computed \emph{proximity graphs}\cite{Simas:2015}, $P_w(X)$ for each time-window resolution $w$. Thus, the entries of the adjacency matrix $P_w$ of a proximity graph are given by:

\begin{equation}
\label{eq:prx_jac}
p_{ij} = \frac{r_{ij}}{r_{ii}+r_{jj}-r_{ij}},\quad \forall_{x_i,x_j \in X}
\end{equation}

\noindent where $p_{ij} \in [0,1]$ and $p_{ii}=1$; $p_{ij}=0$ for terms $x_i$ and $x_j$ that never co-occur in the same time-window in any timeline,
and $p_{ij}=1$ when they always co-occur.
%in the same time-window of every timeline either of them occurs in.
%
This measure is the probability that two terms are mentioned in the same time window, given that one of them was mentioned\cite{Rocha:MyLibraryLANL:2005,Simas:2015}.
To ensure enough support exists in the data for proximity associations, we computed proximity weights only when $r_{ii}+r_{jj}-r_{ij} \geq 10$;
%
%This means that we only consider the proximity between terms $x_i$ and $x_j$, if jointly they occur in at least 10 time windows in all timelines;
%
if $r_{ii}+r_{jj}-r_{ij} < 10$, we set $p_{ij}=0$.

Proximity graphs are \textit{associative knowledge networks}.  %that represent how often items co-occur in a large set of observation units such as documents or, in the present work, time-windows in Instagram timelines.
As in any other co-occurrence method, the assumption is that items that frequently co-occur are associated with a common phenomenon. These knowledge networks have been used successfully for automated fact-checking\cite{Ciampaglia:2015}, protein-protein interaction extraction %\cite{Verspoor:2005,Abi-Haidar:2008}
\cite{Abi-Haidar:2008}, and recommender systems \cite{Rocha:MyLibraryLANL:2005,Simas:2015}.
Here we use them to reveal strong associations of DDI-related terms
%, which may be useful
for public health monitoring.
%
%
%Even though it is not our focus here, we should note that the same similarity could be calculated between user timelines $U$ which can be found useful for detecting users that have similar term use and can then be analysed in greater detail by specialists (eg: users that consumed cannabis, fluoxetine and also had migraine).
%
%In addition to proximity,
%Many network analysis methods, such as those which depend on shortest-path calculations, rely on the isomorphic concept of distance \cite{Simas:2015}. Thus,
We also compute distance graphs $D_w(X)$ for the same time-window resolutions, using the map:

\begin{equation}
d_{ij} = \frac{1}{p_{ij}} - 1
\label{eq:prox2dist}
\end{equation}

In some of our analysis below, we compute the metric closure $D^{C}_w (X)$ of the distance graphs, which is isomorphic to a specific transitive closure of the proximity graph\cite{Simas:2015}.
The metric closure is equivalent to computing the shortest paths between every pair of nodes in the distance graph. Thus, $d^{C}_{ij}$ is the length (sum of distance edge weights) of the shortest path between terms $x_i$ and $x_j$ in the original distance graph $D_w(X)$,
% RBC: Added line (PSB-Reviewer 3)
%All algorithm runs were performed on a personal computer summing a runtime of less than 10 min and they have been proved to
and is known to scale well \cite{Ciampaglia:2015}.
%
%Interestingly, there is an invariant subgraph of $D_w(X)$ when computing the metric closure which is called the \emph{metric backbone}\cite{Simas_rocha:2015}. In other words, some edges $d_{ij}$ in $D_w(X)$ do not change their distance weight when computing shortest paths, because the there is no shorter indirect distance via other nodes in the graph, therefore $d_{ij} = d^{C}_{ij}$; these are \emph{metric edges} because they obbey the triangle inequality. However, there are also \emph{semi-metric edges} which break the triangle inequality, whereby there exist shorter indirect paths than the direct distance: $d_{ij} > d^{C}_{ij}$ \cite{Rocha:2002,Rocha:MyLibraryLANL:2005,Simas:2015,Simas_rocha:2015}.

%To compute the degree of semi-metricity of the edge $d_{ij}$ between nodes/terms $x_i$ and $x_j$ we employ two measures, $s_{ij}$ and $b_{ij}$\cite{Rocha:MyLibraryLANL:2005}:
%
%\begin{equation}
%s_{ij} = \frac{d_{ij}}{d^{C}_{ij}} \quad , \quad b_{ij} = \frac{\langle d_{i}\rangle }{d^{C}_{ij}}  \quad \forall_{x_i,x_j \in X}
%\label{eq:s_b}
%\end{equation}
%
%\noindent where $\langle d_i \rangle$ is the mean direct distance from $x_i$ to all other $x_k \in X$ such that $d_{ik}$ is finite.
%%
%$s_{ij} >1$ for semi-metric edges, and 1 otherwise.
%%
%$b_{ij}$ is only computed for edges that do not exist originally in $D_w(X)$ (i.e. $d_{ij} = \infty$), and it measures how much the shortest indirect distance between $x_i$ and $x_j$ falls below the average distance of $x_i$ to all its directly linked nodes $x_k$. Note that $b_{ij} \neq b_{ji}$.

%
% Results
%
%\section{Results}
%\label{results}

\begin{figure}[h]
\centering
\includegraphics[resolution=300,width=4.5in]{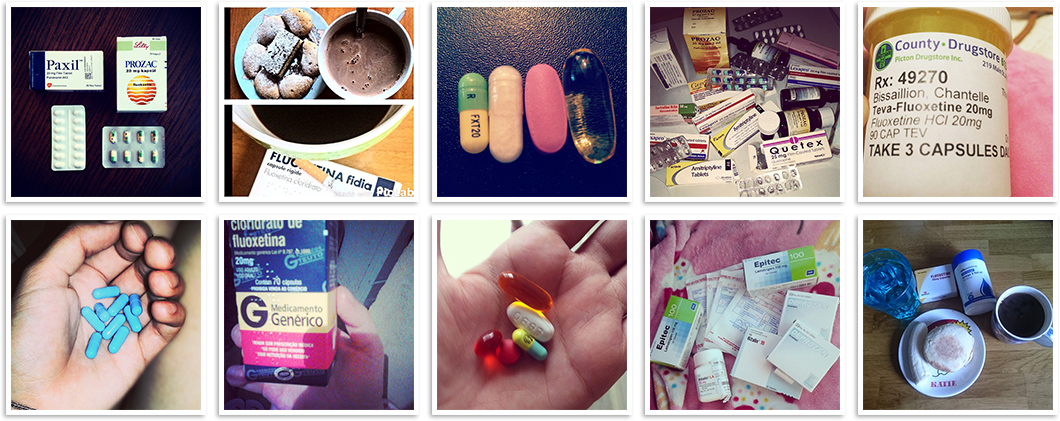}
\caption{Sample of images from collected posts related to \texttt{fluoxetine}.}
\label{fig:sample}
\end{figure}

\section{A Monitoring tool for user-level behavior}
\label{Sec:userLevelTool}

From the analysis of user timelines, it is clear that \emph{Instagram} is a social media platform with much data relevant for public-health monitoring.
Users often discuss personal health-related information such as diagnoses and drugs prescribed.
Photos posted (e.g. Figure \ref{fig:sample}) often depict pills and packaging, along with discussions of intake schedules, expectations and feelings.
%, which can be used to validate the intent and mood of the user in downstream analysis
%

\begin{itemlist}
\item User A on May 25, 2014:
	\begin{unnumlist}
	\item \scriptsize{``\#notmypic .. Say hello to my new friend! Fluoxetina! Side effects by now are a bit of nausea and inquietude.. Better than zoloft! Yesterday night i started to cry while i was with my 2 friends because my ex, bulimia's stress.. I'm sure they thought i'm crazy so i felt like i had to explain my reasons with one of those friends.. Now i'm terrified of his reaction, he is even a friend of my ex.. Don't know what to expect.. It's so hard telling someone about ED and bulimia 😔😔. I'm also thinking about a b/p session today after 2 days clean, maybe it's not the right solution. Idk. \#bulimia \#bulimic \#mia \#ed \#edfamily \#eatingdisorder \#prorecovery \#bingepurge \#purge \#binge \#fat \#prozac \#fluoxetine \#depression \#meds''}
	\end{unnumlist}

\item User B on May 13, 2015:
	\begin{unnumlist}
	\item \scriptsize{``I start fluoxetine tomorrow, the doctor switched me from citalopram to this so let's hope it goes better this time \#anxietymeds \#depressionmeds \#citalopram \#fluoxetine \#anxiety \#depression''}
	\end{unnumlist}
	
\item User B on May 14, 2015 (one day later):
	\begin{unnumlist}
	\item \scriptsize{``ok so I don't know if it's the tablets that are doing this but I feel the lowest I've ever felt and I'm hoping it's not the tablets. Hopefully it's just a bad day, not that there are many good days😢 I hope tomorrow is a better day for everyone, especially if you are feeling the same way I am. \#fluoxetine \#depression \#anxiety \#depressionmeds \#anxietymeds''}
	\end{unnumlist}

\item User C on Feb 05 2014:
	\begin{unnumlist}
	\item \scriptsize{``i survived another trip to the clinic, saw a specialist, did a test that explained i'm an INFJ (introvert) which is apperently only 1\% of the population. Added risperidone and upped ritalin as well as prozac. considering this keeps me `sane' and able to assimilate into the chaos of everyday life i think this counts as my \#100happydays today \#findhappinessineachday \#bipolar \#borderlinepersonalitydisorder \#INFJ \#manicdepression \#goinggovernment \#prozac \#lamotragine \#ritalin \#risperidone''}
	\end{unnumlist}
\end{itemlist}

%User A describes in one post two side effects she is having with the introduction of fluoxetine: nausea and inquietude. He/She even writes about zoloft, the old medication. User B displays temporal information: on May 13 he/she writes about starting on flouxetine and then, on the next day, he/she complains about ``feeling the lowest ever'', possibly a side effect. Moreover, User C describes the prescription switch from  citalopram to fluoxetine adding comments on her doctor's diagnose. These are just a few examples of the richness of the dataset which along with descriptive captions and pictures also contains comments from other users: not focus on this work.

%
% 1) the development of a monitoring tool to easily observe user-level timelines associated with drug and symptom terms of interest,
%

\begin{figure}[h]
\centering
\includegraphics[resolution=300,width=4.5in]{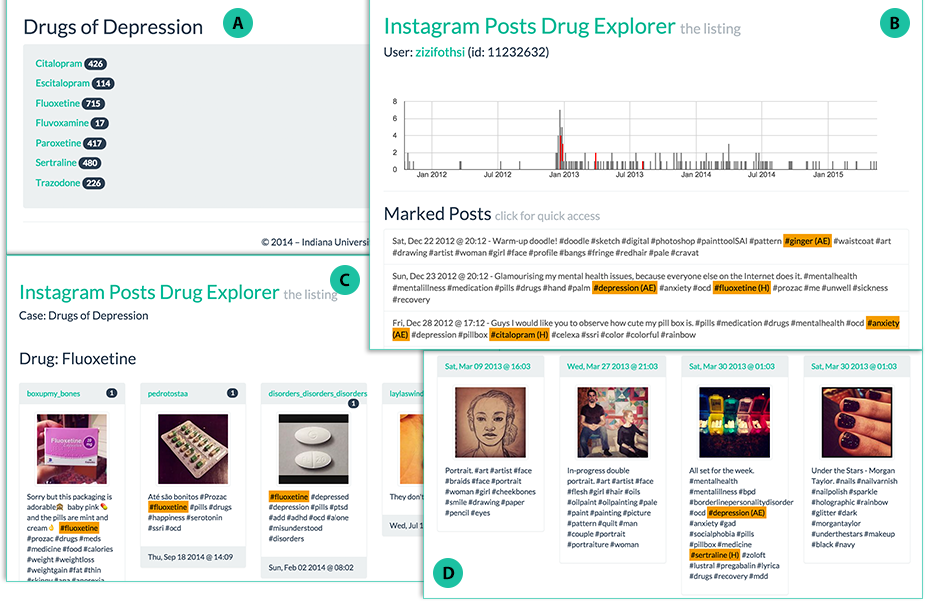}
\caption{Instagram Drug Explorer. See text for explanation.}
\label{fig:explorer}
\end{figure}

\begin{figure}[h]
\centering
\includegraphics[resolution=300,width=4in]{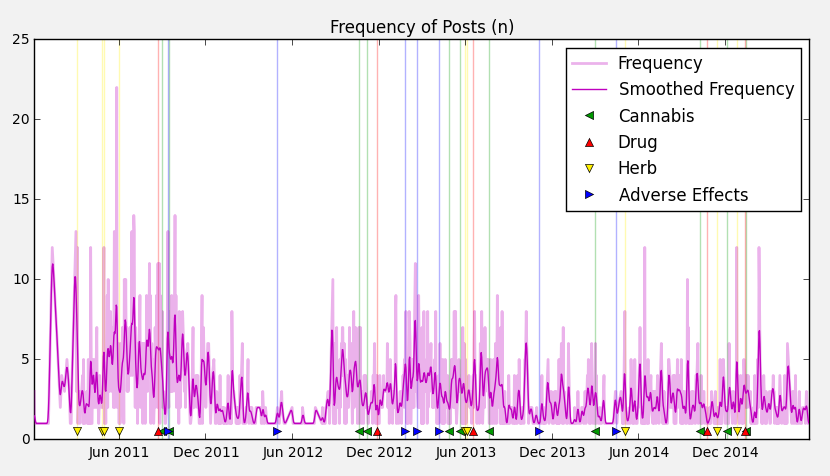}
\caption{User timeline showing daily frequency of posts in time; dictionary terms from are tagged in time. }
\label{fig:ts}
\end{figure}

Given the rich data users post on \emph{Instagram}, from the perspective of public-health monitoring, it is useful to be able to quickly navigate and extract posts and user timelines associated with drugs and symptoms of interest.
For that purpose, we developed the \textit{Instagram Drug Explorer}\footnote{\texttt{http://informatics.indiana.edu/rocha/IDE}.}, a web application to explore, tag, and visualize the data.
This tool also allows downstream improvement of our dictionaries by observing important discourse features not tagged.
Figure \ref{fig:explorer} shows four screenshots with some of the current features: A) the possibility of defining multiple drugs of interest per project; B) a user timeline view that tags class-specific dictionary matches and displays post frequency in time and where individual posts can be quickly selected to be C) visualized separately; D) a summary of posts from user timelines of interest.
Another feature (not shown) is the display of geo-located posts using overlay maps, which can be useful, for instance, to monitor users in places of interest, such as schools, clinics, and hospitals.
%
%LMR: Make sure this is done by conference
%
Using this tool to inspect and select timelines with high number of matches, we were able to identify particularly relevant user timelines such as the one depicted in Figure \ref{fig:ts}, which contains matches from all four dictionaries, and varying post frequency.
%
%LMR: we should see if there are correlations between post frequency and adverse effects and drugs

% 2) population-level behavior via the analysis of co-occurrence networks computed from user timelines at three different scales: monthly, weekly, and daily occurrences.
%
\section{Network analysis of associations in population-level behavior}
\label{Sec:PopBehavior}

Using the proximity or the isomorphic distance graphs (\S \ref{Sec:Data_Methods}), we can explore strong pairwise term associations that arise from the collection of $5,329,720$ posts from the population of $6,927$ users in the study.
The assumption is that dictionary terms that tend to co-occur in a substantial number of user timelines may reveal important interactions among drugs, symptoms, and natural products.
Moreover, because we computed these knowledge networks at different time resolutions, we can explore term associations at different time scales: day, week, and month.
Naturally, a statistical term correlation is not necessarily a causal interaction; also a drug-symptom association may reveal a condition treated by the drug, rather than an adverse reaction.
But large-scale analysis of social media data for relational inference must start with the identification of multivariate correlations, which can be subsequently refined, namely with supervised classification and NLP methods.
Here, as a first step in the analysis of \emph{Instagram} data for public health monitoring, we use unsupervised network science methods to extract term associations of potential interest.

%
%Our belief is that if two term are mentioned in the same timeline within the time resolution there is a possibility they relate to each other. For example, if on the same week User E mentions the words ``fluoxetine'', ``anxiety'' and ``insomnia'', we believe there might be a possible ADR between ``fluoxetine'' and the symptoms ``anxiety'' and ``insomnia'' -- thus we add edges between the user and those terms. The more users mention the same terms, the higher our belief they relate. We also assume the same for a possible DDI if User F mentions, for example, ``fluoxetine'' and ``citalopram'' within the same resolution. Also, by using three different resolution we aim at capturing symptoms or interaction that occur in different time scales. For example, the user may express the adverse effects for ``citalopram'' only a few days after the initial posting of the start of the treatment. This will possibly allows us to map connections that will only be active in a greater (or finer) resolution, on the other hand it will possibly increase the amount of noise in the results.

%From the point of view of an public health expert, one might be interested in those terms for which we have a direct association and are thus possible drug-symptoms associations. From the proximity network $P$ we can then ask which are the edges whose values are higher than a certain threshold. This will result in a network of direct evidence of association.

\begin{figure}[h]
\centering
\includegraphics[resolution=300,width=4.5in]{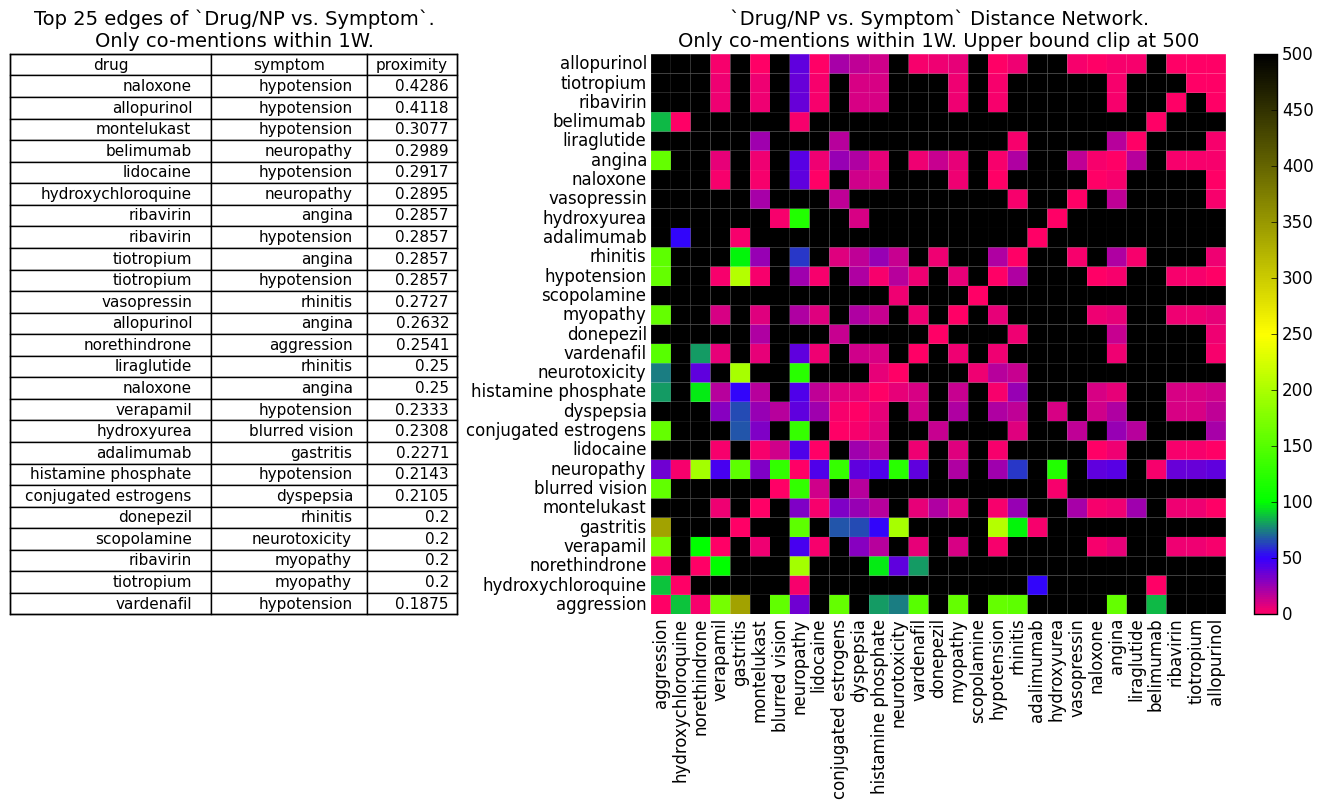}
\caption{drug/NP vs symptom subnetwork: (left) Top 25 pairs with largest proximity correlation. (right) adjacency matrix of distance subnetwork; nearest (furthest) term pairs in red (black).}
\label{fig:distance}
\end{figure}

\begin{figure}[h]
\centering
\includegraphics[resolution=300,width=\linewidth]{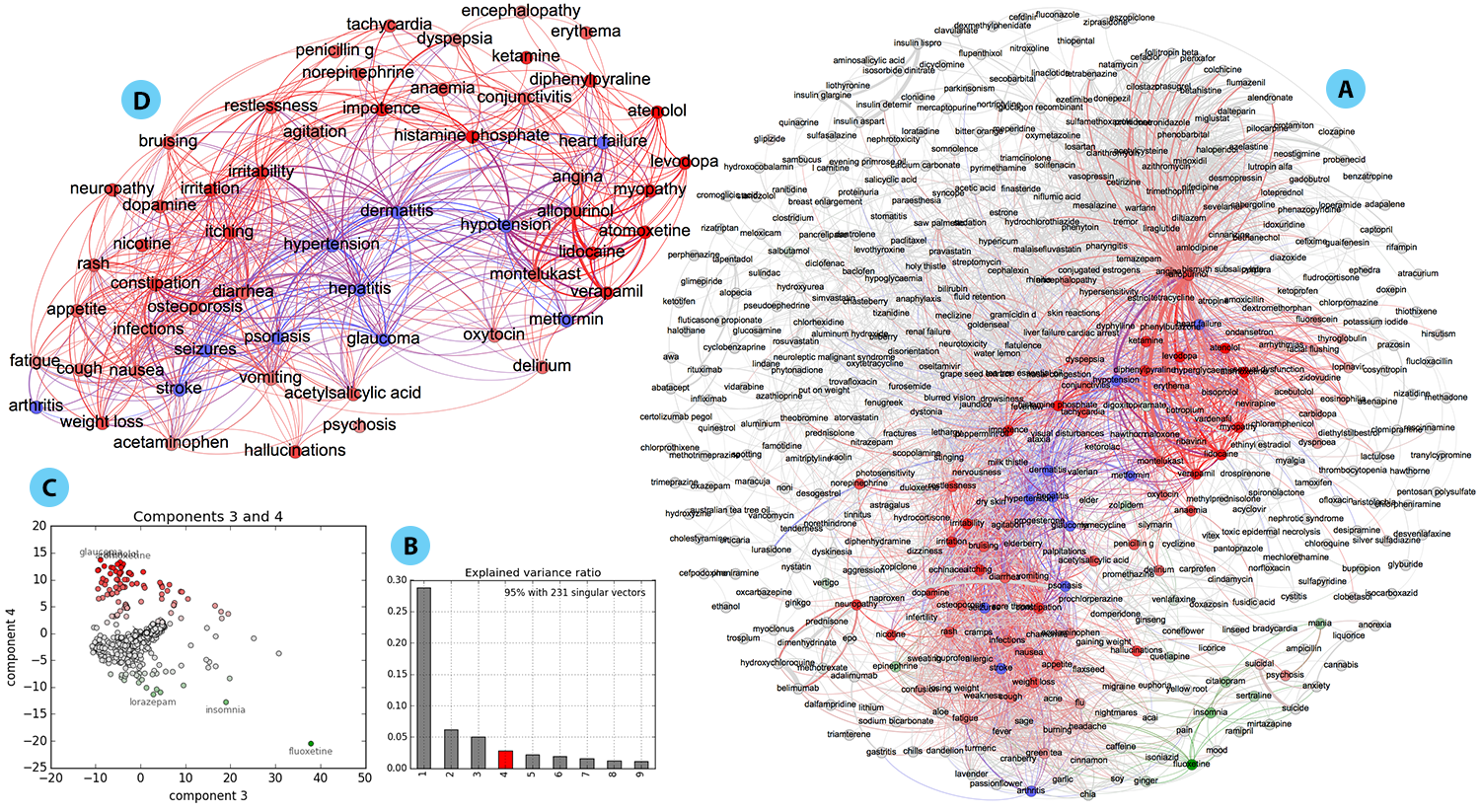}
\caption{A. Largest connected component of the proximity network for 1 week time resolution; weights shown only for $p_{ij} \geq 0.05$ with unconnected terms removed. Edges are colored according to correlation with PC 4. B. Spectrum of the PCA of the proximity network adjacency matrix. C. Biplot of correlation of terms with PC 3 and 4; red (green) terms are most (anti-) correlated with PC4. D. Subgraph depicting the network of terms most correlated with PC4, which is related to \texttt{Psoriasis}; blue nodes depict conditions linked to this complex disease (see text for details); weights shown only for $p_{ij} \geq 0.05$.}
\label{fig:psoriasis_network}
\end{figure}

Consider the proximity networks $P_w(X)$
%of terms $X$
for time resolution $w=1$ week. The full network contains $|X| = 636$ terms (see Figure \ref{fig:psoriasis_network}A for its largest connected component);
Figure \ref{fig:distance} (left) lists the top 25 drug/NP vs symptom associations, as well as the adjacency matrix of the distance subgraph $D_w(X)$ for these drug/NP and symptom pairs (right).
The proximity and distance graphs are isomorphic (\S \ref{Sec:Data_Methods}), but proximity edge weights (left) are directly interpretable as a co-occurrence probability (eq. \ref{eq:prx_jac}), while the isomorphic nonlinear map to distance (eq. \ref{eq:prox2dist}) provides greater discrimination in the visualization of the adjacency matrix (right).
%RBC: In the above paragraph, I've changed  "nolinear" to "nonlinear"

Of the 25 to associations listed in Figure \ref{fig:distance} (left), 12 are known or very likely ADR, 7 do not have conclusive studies but are deemed possible ADR from patient reports, 4 refer to associations between drugs/NP and symptoms they are indicated to treat, 1 has been shown to not be ADR,  and 1 is unknown  (evidence in SI).
Thus, the strongest edges in the 1 week resolution network are relevant drug/NP-symptom associations.
Furthermore, our methodology allows an analyst to collect (via the Drug Explorer tool \S \ref{Sec:userLevelTool}) all the individual timelines and posts that support every association (edge) in the proximity networks, supporting a much more detailed study of the affected population---including for the purpose of fine-tuning dictionaries and mining techniques to better capture the semantics of specific populations.

%

%%%

The proximity networks $P_w(X)$
%of terms $X$  at a given time resolution $w$
also allow us to visualize, explore and search the ``conceptual space'' of drugs, symptoms, and NP as they co-occur in the depression timeline dataset.
The largest connected component of the proximity network for $w=1$ week is shown in Figure \ref{fig:psoriasis_network}A.
The network representation allows us to find clusters of associations, beyond term pairs, which may be related via the same underlying phenomenon. Many multivariate and network analysis methods can be used to uncover modular organization \cite{fortunato2010community}. To exemplify, here we use the Principal Component Analysis (PCA) \cite{wall2003singular} of the proximity network adjacency matrix, which reveals potential phenomena of interest.
%in the depression timeline dataset.

%RBC: Psoriasis is not the MOST correlated term now. It's the 7th. (Glaucoma: 1st; Atenolol: 2nd; Atomoxetine: 3rd; Hepatitis: 4th; Levodopa: 5th; Histamine Phosphate: 6th)
For instance, Figure \ref{fig:psoriasis_network},
%we can see that component 4 is associated with \texttt{Psoriasis}. More specifically, t
depicts a set of terms correlated with principal component (PC) 4 (red)---others could be chosen (see SI).
%
%with \texttt{Psoriasis} being the 7th term most correlated.
The subnetwork of these terms is depicted in Figure \ref{fig:psoriasis_network}D.
and it reveals a set of terms denoting a complex interaction of conditions which are coherent with what is becoming known about \texttt{Psoriasis}.
Several of the edges associate terms related to heart disease, stroke, hypertension, hypotension, and diabetes which are high risks for \texttt{Psoriasis} patients\cite{WebMD:PsoriasisRisk}, including potential drug interactions (\texttt{Metformin} for Diabetes, \texttt{Verapimil} for high blood pressure and Stroke).
This subnetwork also reveals associations with \texttt{Psoriasis} which are currently receiving some attention, such as with viral hepatitis\cite{cohen2010psoriasis} and seizure disorder\cite{Mandl_epilepsy_sautoimmune}.
Naturally, the network also includes many terms associated with skin infections and immune reactions.
The \texttt{Psoriasis}  subnetwork is just an example of a multi-term phenomenon of interest that is represented in the whole network; other PCA components are shown in SI, including additional analysis of the Psoriasis subnetwork. Importantly, we can identify users who may be experiencing this cluster of symptoms by following the posts and timelines behind the weights in the subnetwork, which is useful for public health monitoring.

While the \texttt{Psoriasis} subnetwork was discovered purely by data-driven analysis, another way to use these networks is to to query them for specific terms most associated with a set of drugs or symptoms of interest.
This problem of finding which other items $A \subseteq X$ are near a set of query items $Q \subseteq X$ (including a subnetwork of interest) is common in recommender systems and information retrieval\cite{Rocha:MyLibraryLANL:2005}.
The answer set $A$ can be computed as:

\begin{equation}
A \equiv \Bigg\{ x_j :  \forall_{x_i \in Q} \quad \underset{x_j \in X-Q }{\Phi} ( p_{ij}) \geq \alpha \Bigg\}
\label{eq:query}\\
\end{equation}

\noindent where $\Phi$ is an operator of choice, $p_{ij}$ is the proximity weight between terms $x_i$ and $x_j$ (\S \ref{Sec:Data_Methods}), and $\alpha$ is a desired threshold.
If we are interested in a set of terms $A$ which are strongly related to \emph{every} term in query set $Q$, then we use $\Phi = \min$. If we are interested in terms strongly related to \emph{at least one} term in $Q$, then $\Phi = \max$. For a compromise between the two, we can use $\Phi =$ avg (average).
%RBC: These terms were updated (after caffeine bug)
Consider the query $Q = \{\texttt{fluoxetine}, \texttt{anorexia}\}$ on the network of Figure \ref{fig:psoriasis_network}A ($w=1$ week). Using  $\Phi = \min$, we obtain an answer set with terms strongly related to both query terms (ordered by relevance): $A = \{ \texttt{suicidal}$, \texttt{suicide}, \texttt{anxiety}, \texttt{pain}, \texttt{mood}, \texttt{cinnamon}, \texttt{insomnia}, \texttt{soy}, \texttt{headache}, \texttt{mania}, \texttt{chia}, \texttt{cannabis} $\}$.
%RBC: These terms were updated (after caffeine bug)
For the query $Q = \{ \texttt{psoriasis}$, \texttt{heart failure}, \texttt{stroke} $\}$ using $\Phi = $ avg, we obtain (ordered by relevance): $A = \{ \texttt{infections}$, \texttt{diarrhea}, \texttt{hypertension}, \texttt{seizures}, \texttt{hepatitis}, \texttt{constipation}, \texttt{dermatitis}, \texttt{glaucoma}, \texttt{vomiting} $\}$, which relates to the discussion above.
Additional query examples and details of the network search interface are shown in SI.
%

%
% In our results, if we query the monthly resolution network for the terms ``cannabis'' and ``pain'' our results are (mean proximity shown): ``caffeine'' (.1113), ``anxiety'' (.1039), ``mood'' (.1005), ``insomnia'' (.0877), ``fluoxetine'' (.0847). Interestinly, if we query for ``hepatitis'', the results are ``impotence'' (.15), ``hypertension'' (.147), ``seizures'' (.1263), ``osteoporosis'' (.1194), ``infections'' (.1181) and ``neurotoxicity'' (.1176).

%Figure \ref{fig:query} shows the output of the network search interface

%
% Note that ``anxiety'' (.0934) and ``pain'' (.0327) are the top results.
%
%\begin{figure}[h]
%\centerline{\psfig{file=../images/query,width=3.2in}}
%\caption{Network query for terms ``flouxetine'' and ``anorexia'' ordered by the $min$ operator.}
%\label{fig:query}
%\end{figure}

%
% 3) Analysis of these networks further reveals 3) drug and symptom direct and indirect associations with greater support in user timelines
%

%

%\section{Network analysis of indirect associations in population-level behavior}
%\label{Sec:PopBehaviorIndirect}

Proximity $P_w(X)$ networks are useful to discover associations between terms which co-occur in time windows $w$ of user timelines (\S \ref{Sec:PopBehavior}).
But they are also useful to infer \emph{indirect associations} between terms. In other words, terms that do not co-occur much in user timelines, but which tend to co-occur with the same other terms.
In network science
%we think of these as \emph{semi-metric} associations (\S \ref{Sec:Data_Methods}), which
%
indirect associations are typically obtained via the computation of shortest path algorithms on the isomorphic distance graphs $D_w(X)$ \cite{Simas:2015}.
Terms which are very strongly connected via indirect paths, but weakly connected via direct edges, break
%the triangular inequality (or a generalized measure of transitivity).
transitivity criteria\cite{Simas:2015}.
%
%The assumption is that such semi-metric or non-transitive associations carry \emph{latent} information not directly observed in the data, but which is strongly implied by the network via indirect paths.
%
We have previously shown that such indirect paths are useful to predict novel trends in recommender systems
%\cite{Rocha:2002,Simas:2015}
\cite{Simas:2015}, and are also instrumental to infer factual associations in knowledge networks\cite{Ciampaglia:2015}.
In this context, the hypothesis is that strongly indirectly associated terms may reveal unknown DDI and ADR.

%\begin{figure}[h]
%\centerline{\psfig{file=../images/1W/img-instagram-closure-m-w-custom.png,width=5in}}
%\caption{``drug'' vs ``symptom'' metric closure subnetwork. Edges where association distance became shorter after closure: (left) Top 25 semi-metric edges. (right) subnetwork heatmap. Edges in this network}
%\label{fig:closure}
%\end{figure}

\begin{figure}[h]
\centering
\includegraphics[resolution=300,width=4.5in]{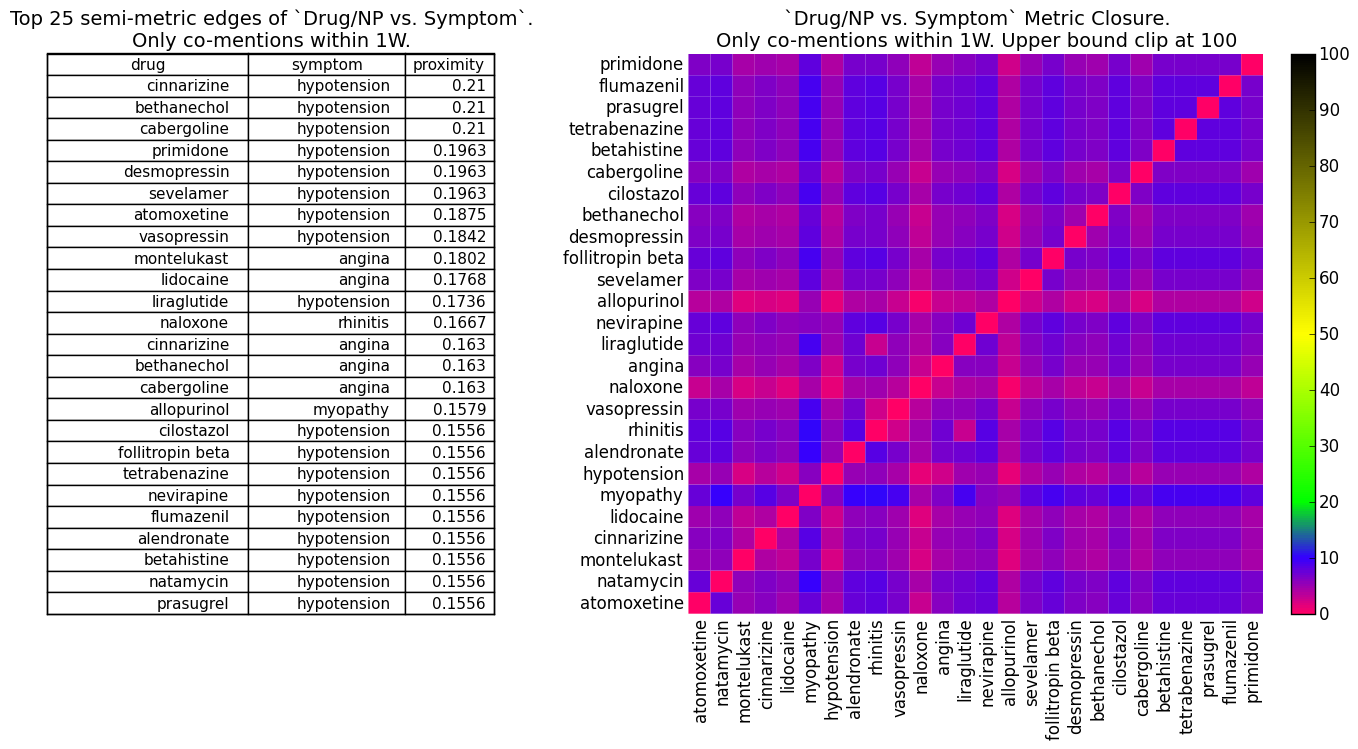}
\caption{drug/NP vs symptom subnetwork after shortest path calculation. (left) Top 25 non-transitive term pairs. (right) adjacency matrix of distance subnetwork after shortest path calculation.}
\label{fig:closure}
\end{figure}

To find the term pairs that most break transitivity we compute all shortest paths in the networks (via Dijkstra's algorithm): the metric closure $D_w^C(X)$.
Figure \ref{fig:closure} lists the top 25 drug/NP vs symptom associations which most break transitivity. In other words, these are term pairs which are very strongly associated via indirect paths, but very weakly associated directly.
Of the extracted associations listed in the table of Figure \ref{fig:closure}, 6 are known or likely ADR, 3 are possible ADR from patient reports but no conclusive study, 2 refer to associations between drugs/NP and symptoms they are indicated to treat, and all other 14 are unknown  (evidence provided in SI).
Thus, unlike the case of direct associations (Figure \ref{fig:distance}), there is less evidence for the indirect associations in the literature. This could be because they are false associations, or because they have not been discovered yet. Validating these associations empirically is left for forthcoming work; here the goal is to show how network analysis methods can be used to select such latent associations which are highly implied by indirect paths (transitivity) but are not directly observed in user post co-mentions.

Similarly to what was done with direct associations above, we can also query the proximity network obtained after shortest path computation $P_w^C(X)$ (the isomorphic  proximity graph to $D_w^C(X)$ via eq. \ref{eq:prox2dist}).
For instance, if we query the original $w = 1$ week proximity network $P_w^C(X)$ (the one depicted in Figure \ref{fig:psoriasis_network}A) with
$Q = \{\texttt{psoriasis}, \texttt{metformin}\}$ (a type 2 diabetes drug), using  $\Phi = \min$, we obtain $A = \{\texttt{montelukast}$ , \texttt{hypertension}, \texttt{dermatitis}, \texttt{hypotension}, \texttt{hepatitis}$\}$ as the top 5 terms---\texttt{montelukast} is a drug used to treat allergies.
If we now use the same query $Q$ on the metric closure network $P_w^C(X)$ instead, the top 5 answer set becomes $A^C = \{\texttt{montelukast}$, \texttt{hypotension,} \texttt{naloxone}, \texttt{allopurinol}, \texttt{hypertension}$\}$  (full query results in SI).
%
%RBC: (PSB: Reviewer 2) = naloxone is a "synthetic opiate antagonist" not a "narcotic"
In other words, after computing shortest paths,\texttt{naloxone} (a synthetic opiate antagonist used to reverse the effects, including addiction, caused by narcorics) and \texttt{allopurinol} (a drug used to treat gout, kidney stones, and decrease levels of uric acid in cancer patients), become more strongly associated with the query terms.
These indirect associations to do not occur very strongly in the observed \emph{Instagram} timeline data, but are strongly implied by indirect paths in the network of term proximity.
In this case, the \emph{latent} associations may provide additional evidence supporting recent observations that psoriasis (an autoimmune condition) is linked to heart disease, cancer, diabetes and depression\cite{WebMD:PsoriasisRisk}.

\section{Discussion and Future Directions}

Our preliminary analysis demonstrates that there exists a substantial  health-related user community in \textit{Instagram} who posts about their health conditions and medications. The drug, NP and symptom dictionaries we employed extracted a large number of posts with such data, enough to build knowledge networks of hundreds of terms representing the pharmacology and symptomatic ``conceptual space'' of \emph{Instagram} users posting about depression.
Our results and software further demonstrate that such space can be navigated for public health monitoring, whereby analysts can search and visualize user timelines of interest. Furthermore, the network representation of this space allows us to extract population-level term associations and subnetworks of terms arising from underlying (modular) phenomena of interest---such as the Psoriasis network involving various related conditions.
Thus, \textit{Instagram} data shows great potential for public health monitoring and surveillance for DDI and ADR.

Direct associations in the knowledge networks are substantiated by actual co-mentions in posts from user timelines, which can subsequently be retrieved by public health analysts using our drug explorer application. In our preliminary work, the top extracted direct associations are shown to be backed by the literature, but we intend to pursue the systematic validation of such associations in future work. 
%
%Our extracted networks will be made available (upon publication) to the community interested in public health and biocomputing, so we hope that other groups may participate in the validation of this data.
%
Network methods also allow us to uncover indirect associations among terms. These may reveal latent, yet unknown, associations, and as such, very relevant for public health monitoring. Studying the network of indirect associations can be further used to understand community structure as well as redundancy in the data, which we intend to study next.

We have analyzed posts and user timelines related to depression only. Adding additional conditions of interest (e.g. epilepsy or psoriasis) to extract additional posts would monitor different communities, and would likely improve the overall extraction of associations, which we intend to test in the near future.
While the drug dictionary is quite well developed already, the NP and symptoms dictionaries need to be further developed, especially towards increasing the terminology associated with symptoms as well as on catching particular linguistic expressions of symptoms in Instagram. The development of named entity recognition tailored to Instagram is another avenue we intend to pursue,  starting from and expanding what has already been done for Twitter\cite{gonzalez2015}.

The methodology we describe here allows us to discern drug, NP and symptom associations derived from user timeline co-mentions at different timescales. All the results displayed pertain to a one week window, however we also computed day and month windows. The comparison of results at different timescales would allow, in principle, the discovery of more immediate as well as more delayed interactions. Such a comparison is also something we intend to pursue in forthcoming work.
Finally, the timeseries analysis of user timelines can be used to detect discernible changes in behavior for users and groups of users. One could track, for instance, critical changes in mood associated with the onset of depression \cite{van2014critical}, which constitutes yet another exciting avenue to pursue with this line of research.

Our preliminary analysis demonstrates that \emph{Instagram} is a very powerful source of data of potential benefit to monitor and uncover DDI and ADR. Moreover, our work shows that complex network analysis provides an important toolbox to extract health-related associations and their support from large-scale social media data.

%
% Acknowledgments
%
\section*{Acknowledgments}

%%% VERIFICAR SE O NUMERO DO NIH ESTA CORRETO
%%% ACHEI DOIS NUMEROS DIFERENTES NO SITE DA NIH
%%% 1R01LM011945-01 === http://projectreporter.nih.gov/project_info_details.cfm?aid=8686117&icde=25589859
%%% e
%%% 5R01GM104483-02 === http://projectreporter.nih.gov/project_info_description.cfm?aid=8913218&icde=25589859

This work was supported by a grant from the National Institutes of Health, National Library of Medicine Program, grant 01LM011945-01``BLR: Evidence-based Drug-Interaction Discovery: In-Vivo, In-Vitro and Clinical,'' and a grant from Persistent Systems. RBC is supported by CAPES Foundation Grant No. 18668127. The funders had no role in study design, data collection and analysis, decision to publish, or preparation of the manuscript.

%
% References
%
\bibliographystyle{ws-procs11x85}
\bibliography{references}

% Include SI
\setboolean{@twoside}{false}


\section*{Supplementary material}

# Supplemental Information

*Authors: Rion Brattig Correia, Lang Li & Luis M. Rocha\**

*\*rocha@indiana.edu*

## Hashtags collected

These are the drug names that were collected using the Instagram API. A post was collected if the user mentioned the name of the drug -- or its synonyms -- as a hashtag (#) in the post caption or comment with it.

| Drug | # Posts | Synonyms tags |
| --- | --- | --- |
| sertraline | 574 | sertralina |
| fluoxetine | 8143 | fluoxetin, fluoxetina, fluoxetinum, fluoxétine, prozac |
| citalopram | 426 | citadur, nitalapram |
| escitalopram | 117 | escitalopramum, esertia |
| paroxetine | 470 | paroxetina, paroxetinum |
| fluvoxamine | 22 | fluvoxamina, fluvoxaminum |
| trazodone | 227 | trazodona, trazodonum |

## Herbal Medicine Terms

The herbal medicine terms were extracted from the US National Library of Medicine. For each herb all it's common names were included, for example: "açaí" (herb) and "amazonian palm berry" (synonym).
reference: http://www.nlm.nih.gov/medlineplus/herbalmedicine.html

acai, amazonian palm berry, aloe vera, aloe, burn plant, lily of the desert, elephants gall, aristolochic acids, aristolochia, asarum, asian ginseng, ginseng, chinese ginseng, korean ginseng, asiatic ginseng, astragalus, bei qi, huang qi, ogi, hwanggi, milk vetch, bilberry, european blueberry, whortleberry, huckleberry, bitter orange, seville orange, sour orange, zhi shi, black cohosh, black snakeroot, macrotys, bugbane, bugwort, rattleroot, rattleweed, butterbur, petasites, purple butterbur, petadolex, cats claw, una de gato, chamomile, german chamomile, chasteberry, chaste-tree berry, vitex, monks pepper, cinnamon, cinnamon bark, ceylon cinnamon, cassia cinnamon, chinese cinnamon, cranberry, american cranberry, bog cranberry, dandelion, lions tooth, blowball, echinacea, purple coneflower, coneflower, american coneflower, ephedra, chinese ephedra, ma huang, elderberry, european elder,  black elder, elder, elderberry, elder flower, sambucus, mistletoe, european mistletoe, epo, evening primrose oil, fenugreek, fenugreek seed, feverfew, bachelors buttons, featherfew, flaxseed, flaxseed oil, linseed, garlic, ginger, ginkgo, ginkgo biloba, fossil tree, maidenhair tree, japanese silver apricot, baiguo, bai guo ye, kew tree, yinhsing, goldenseal, yellow root, grape seed extract, green tea, chinese tea, japanese tea, hawthorn, english hawthorn, harthorne, hawthorne, haw, hoodia, kalahari cactus, xhoba, horse chestnut, buckeye, spanish chestnut, kava, kava kava, awa, kava pepper, lavender, english lavender, garden lavender, licorice,

licorice root, liquorice, sweet root, gan zao, chinese licorice, milk thistle, mary thistle, holy thistle, silymarin, silybinin, silibinin, silybin, noni, morinda, indian mulberry, hog apple, canary wood, passionflower, maypop, apricot vine, old field apricot, maracuja, water lemon, peppermint oil, red clover, cow clover, meadow clover, wild clover, red yeast rice, sage, black sage, broad-leafed sage, common sage, saw palmetto, american dwarf palm tree, cabbage palm, soy, st. johns wort, hypericum, klamath weed, goatweed, tea tree oil, australian tea tree oil, tea tree essential oil, melaleuca oil, thunder god vine, lei gong teng, turmeric, turmeric root, indian saffron, valerian, all-heal, garden heliotrope, chia, chia seeds, yohimbe, yohimbe bark.

## Cannabis Terms

These are the dictionary of terms for Cannabis used in the analysis:

cannabis, marijuana, marihuana, maryjane, mary jane, doobie, ganja, hashish, bhang, maconha, skunk, weed, dagga, hashish, pot, bud, herb, haze, joint, blunt, chronic, dank, hash, hierba, mota, 420.

## Top 25 Mentions

| Term | # of mentions | Dictionary Type |
|---|---|---|
| cannabis | 66540 | cann |
| anorexia | 26872 | adve |
| anxiety | 26309 | adve |
| pain | 15677 | adve |
| suicide | 11616 | adve |
| mood | 11532 | adve |
| fluoxetine | 9961 | drug |
| suicidal | 8909 | adve |
| ginger | 7289 | herb |
| insomnia | 5917 | adve |
| soy | 4417 | herb |
| garlic | 3804 | herb |
| cinnamon | 3619 | herb |
| caffeine | 2705 | drug |
| burning | 2075 | adve |

| | | |
|---|---|---|
| flu | 2011 | adve |
| lavender | 1860 | herb |
| headache | 1855 | adve |
| chia | 1754 | herb |
| arthritis | 1685 | adve |

Top 25 Mentions in each Dictionary

| Drugs | # posts (# users) | Herbs | # posts (# users) | Adverse Effect | # posts (# users) |
|---|---|---|---|---|---|
| fluoxetine | 9961 | cannabis | 66540 | anorexia | 26872 |
| caffeine | 2705 | ginger | 7289 | anxiety | 26309 |
| sertraline | 1195 | soy | 4417 | pain | 15677 |
| citalopram | 692 | garlic | 3804 | suicide | 11616 |
| epinephrine | 636 | cinnamon | 3619 | mood | 11532 |
| quetiapine | 480 | lavender | 1860 | suicidal | 8909 |
| acetaminophen | 440 | chia | 1754 | insomnia | 5917 |
| lorazepam | 410 | green tea | 957 | burning | 2075 |
| ibuprofen | 370 | cranberry | 784 | flu | 2011 |
| lithium | 306 | turmeric | 771 | headache | 1855 |
| nicotine | 304 | sage | 725 | arthritis | 1685 |
| prednisone | 269 | dandelion | 633 | pancreatitis | 1618 |
| methotrexate | 227 | aloe | 513 | mania | 1533 |
| ethanol | 217 | chamomile | 338 | fever | 1521 |
| acetylsalicylic acid | 212 | flaxseed | 312 | migraine | 1290 |
| cefpodoxime | 197 | acai | 295 | fatigue | 1014 |
| dopamine | 195 | licorice | 191 | psychosis | 961 |
| venlafaxine | 187 | elder | 167 | nightmares | 925 |
| duloxetine | 167 | mistletoe | 157 | cough | 865 |

| | | | | | |
|---|---|---|---|---|---|
| zopiclone | 165 | ginseng | 130 | acne | 807 |
| zolpidem | 157 | passionflower | 113 | weight loss | 753 |
| adalimumab | 152 | liquorice | 97 | confusion | 711 |
| hydroxychloroquine | 127 | echinacea | 95 | weakness | 699 |
| bupropion | 127 | huckleberry | 93 | allergic | 561 |
| diphenhydramine | 121 | kava | 82 | nausea | 556 |

## Statistics of the Network

| Full Network | |
|---|---|
| | 1 Week |
| Edges in D | 34,935 |
| Edges in D^{MC} | 172,578 |
| Metric edges in D | 4,607 (13.19%) |
| Edges with S>1 in D | 30,328 (86.81%) |
| Edges with \inf distance in D | 166,995 |
| Edges with B>1 in D | 130,405 |
| Number of terms: 636 -- (n^2-n)/2  Total number of possible edges: 201,930  Colunas: #edges in original matrix (D), #edges in metric closure (D^mc),  # / % metric edges in original matrix (D), # / % s>1 parameter in original matrix (D), # edges with infinite distance in D, # of edges with b>1 | |

# Preliminary evidence for drug-symptoms associations (Figure 4)

### 1) naloxone - hypotension (proximity: 0.4285) - Symptom

Indications: "For the complete or partial reversal of narcotic depression, including respiratory depression, induced by opioids including natural and synthetic narcotics, propoxyphene, methadone and the narcotic-antagonist analgesics: nalbuphine, pentazocine and butorphanol. It is also indicated for the diagnosis of suspected acute opioid overdose. It may also be used as an adjunctive agent to **increase blood pressure** in the management of septic shock."[1]
Pharmacodynamics: "Naloxone is an opiate antagonist and prevents or reverses the effects of opioids including respiratory depression, sedation and **hypotension**"[1]

references:
1. DrugBank

### 2) allopurinol - hypotension (proximity: 0.4117) -  Possible ADR

Indication: "For the treatment of hyperuricemia associated with primary or secondary gout. Also indicated for the treatment of primary or secondary uric acid nephropathy, with or without the symptoms of gout, as well as chemotherapy-induced hyperuricemia and recurrent renal calculi."[1]
"Hyperuricemia is associated strongly with the development of hypertension, renal disease, and progression. Allopurinol decreases serum uric acid levels by inhibiting the enzyme xanthine oxidase."[2]

Evidence points against it in study of patients who take medication for arterial hypertension [3], however, some evidence for ADR exists from patient reports [4].

references:
1. DrugBank
2. http://www.sciencedirect.com/science/article/pii/S0272638605015180
3. http://www.ncbi.nlm.nih.gov/pubmed/21405957
4. http://factmed.com/study-ALLOPURINOL-causing-HYPOTENSION.php

### 3) montelukast - hypotension (proximity: 0.3076) - Possible ADR

Indications: "For the treatment of asthma"[1]

No study found, but some evidence of ADR especially with patients who take Lorazepam for anxiety and depression [2].

references:
1. DrugBank
2. http://www.ehealthme.com/ds/montelukast+sodium/hypotension

### 4) belimumab - neuropathy (proximity: 0.2989) - Symptom

Indications: "Adjunct treatment for auto-antibody-positive active systemic lupus erythematosus."[1] Systemic lupus erythematosus (SLE) is an autoimmune disease in which the body's immune system mistakenly attacks healthy tissue. It can affect the skin, joints, kidneys, brain, and other organs. Other symptoms depend on which part of the body is affected: **Brain and nervous system**: headaches, **numbness**, tingling, seizures, vision problems, personality changes[2]

references:
1. DrugBank
2. http://www.nlm.nih.gov/medlineplus/ency/article/000435.htm

### 5) lidocaine - hypotension (proximity: 0.2916) - ADR

Indications: "For production of local or regional anesthesia by infiltration techniques such as percutaneous injection and intravenous regional anesthesia by peripheral nerve block techniques such as brachial plexus and intercostal and by central neural techniques such as lumbar and caudal epidural blocks."[1]
"This paper reports the cardiovascular effects of intentionally toxic intravenous doses of **lidocaine** [...] and the mechanisms of death. In 4/4 lidocaine-treated animals, respiratory depression with bradycardia and **hypotension** without arrhythmias were the causes of death."

references:
1. DrugBank
2. http://journals.lww.com/anesthesia-analgesia/Abstract/1989/09000/Myocardial_and_Cerebral_Drug_Concentrations_and.2.aspx

### 6) hydroxychloroquine - neuropathy (proximity: ) - Possible ADR (unknown, 1 case)

Indications: "For the suppressive treatment and treatment of acute attacks of malaria due to Plasmodium vivax, P. malariae, P. ovale, and susceptible strains of P. falciparum. It is also indicated for the treatment of discoid and systemic **lupus erythematosus (LE)**, and rheumatoid arthritis" [1].
Toxicity: "Symptoms of overdose include headache, drowsiness, visual disturbances, cardiovascular collapse, and convulsions, followed by sudden and early respiratory and cardiac arrest. The electrocardiogram may reveal atrial standstill, nodal rhythm, prolonged intraventricular conduction time, and progressive bradycardia leading to ventricular fibrillation and/or arrest."[1]
"Chloroquine and **hydroxychloroquine** (HCQ) are commonly prescribed antimalarial agents used for a variety of systemic diseases.  In this report, we describe **a patient** with rheumatoid arthritis and respiratory failure associated with proximal myopathy secondary to HCQ. Patients treated with HCQ in whom proximal myopathy, **neuropathy**, or cardiomyopathy develop should be evaluated for possible **HCQ toxicity**."[2]

references:
1. DrugBank
2. http://journal.publications.chestnet.org/article.aspx?articleid=1084931

### 7) ribavirin - angina (proximity: 0.2894) - ADR

Indications: "For the treatment of chronic hepatitis C and for respiratory syncytial virus (RSV)."[1]
Contraindications to ribavirin include end-stage renal failure, anaemia, severe heart disease, pregnancy and inadequate contraception. The major side-effect of **ribavirin** is haemolytic anaemia, which can be severe. Cardiovascular disease should be carefully excluded in patients considered for combination therapy, as anaemia may lead to **angina** or **heart failure** in these patients.[2]

references:
1. DrugBank
2. http://ar.iiarjournals.org/content/25/2B/1315.short

### 8) ribavirin - hypotension (proximity: 0.2857) - ADR

Indications: "For the treatment of chronic hepatitis C and for respiratory syncytial virus (RSV)."[1]
"Cardiovascular effects, particularly bradycardia, have been associated with ribavirin use. Bradycardia, hypotension, and cardiac arrest have been reported with inhaled ribavirin"[2]

references:
1. DrugBank
2. http://onlinelibrary.wiley.com/doi/10.1592/phco.27.4.494/abstract

### 9) tiotropium - angina (proximity: 0.2857) - ADR

Indications: "Used in the management of chronic obstructive pulmonary disease (COPD)"[1]
"Cardiac angina was more common on active treatments than placebo"[2]

references:
1. DrugBank

### 10) tiotropium - hypotension (proximity: 0.2857) - Unknown, possible ADR

Indications: "Used in the management of chronic obstructive pulmonary disease (COPD)"[1]. No studies with results found. Some evidence of possible ADR [2].

references:
1. DrugBank
2. http://factmed.com/study-TIOTROPIUM-causing-HYPOTENSION.php

### 11) vasopressin - rhinitis (proximity: 0.2727) - ADR

Indications: "For the treatment of enuresis, polyuria, diabetes insipidus, polydipsia and oesophageal varices with bleeding"[1]
"Also known as "desmopressin" (1-deamino-8-D-argininevasopressin)[2]. Nasal congestion and rhinitis have been reported with nasal spray formulations."[1]
references:
1. DrugBank
2. http://onlinelibrary.wiley.com/doi/10.1046/j.1475-097X.2002.00401.x/full

### 12) allopurinol - angina (proximity: 0.2631) - ADR

Indication: "For the treatment of hyperuricemia associated with primary or secondary gout. Also indicated for the treatment of primary or secondary uric acid nephropathy, with or without the symptoms of gout, as well as chemotherapy-induced hyperuricemia and recurrent renal calculi."[1]
Patients in the **allopurinol** group and in the highest uric acid quartile had indicators of more severe **Heart Failure**[2]
references:
1. DrugBank
2. http://www.sciencedirect.com/science/article/pii/S0002870310007337

### 13) norethindrone-aggression (proximity: 0.2632) - ADR

"Perinatal exposure to **norethindrone** influences morphology and **aggressive** behavior of female mice" [1].
"In the fighting fish Beta splenders it is associated with masculinization, normal growth and secondary sexual characteristics, yet no abnormal aggression behavior" [2].

**references:**
1. http://www.sciencedirect.com/science/article/pii/0018506X81900052
2. http://repository.ias.ac.in/39869/

### 14) liraglutide - rhinitis (proximity: 0.25) - Unknown, possible ADR

Indications: "For use in/treatment of diabetes mellitus type 2."[1]
No studies found, some evidence from patient reports [2]

references:
1. Drugbank
2. http://factmed.com/study-LIRAGLUTIDE%20(NN2211)-causing-RHINITIS%20ALLERGIC.php

### 15) naloxone - angina (proximity: 0.25) - No evidence of ADR

Indications: "For the complete or partial reversal of narcotic depression, including respiratory depression, induced by opioids including natural and synthetic narcotics, propoxyphene, methadone and the narcotic-antagonist analgesics: nalbuphine, pentazocine and butorphanol. It is also indicated for the diagnosis of suspected acute opioid overdose. It may also be used as an adjunctive agent to increase blood pressure in the management of septic shock."[1]

Small study on patients with coronary disease showed no evidence of ADR [2].

references:

1. DrugBank
2. http://www.ncbi.nlm.nih.gov/pubmed/6496361

### 16) verapamil - hypotension (proximity: 0.2333) - ADR

Indications: "For the treatment of **hypertension**, angina, and cluster headache prophylaxis."[1]
references:
The use of **verapamil** in the treatment of cardiac disease is becoming more frequent. In general, it is used for the reversal of a supraver/tricular tachycardia and this can be achieved safely and effectively with the minimum of side effects [4], of which the most common one is transient, mild **hypotension**.

1. DrugBank
2. http://link.springer.com/article/10.1007%2FBF01686855

### 17) hydroxyurea - blurred vision (proximity: 0.2307) - Symptom

Indications: "For management of melanoma, resistant chronic myelocytic **leukemia**, and recurrent, metastatic, or inoperable carcinoma of the ovary and Sickle-cell anemia."[1]
Leukostasis is a white blood cell count above 100,000/μL. It is most often seen in leukemia patients. The brain and lungs are the two most commonly affected organs. Occluded microcirculation causes local hypoxemia and hemorrage manifesting as headache, **blurred vision**.[2]

references:
2. DrugBank
3. https://en.wikipedia.org/wiki/Leukostasis

### 18) adalimumab - gastritis (proximity: 0.2271) - Symptom

"Crohn's disease (CD) is an inflammatory bowel disease. It causes **inflammation of the lining of your digestive tract,** which can lead to abdominal pain, severe diarrhea, fatigue, weight loss and malnutrition"[2]. "CD most often manifests in late adolescence or early adulthood. **Adalimumab** is a fully human immunoglobulin G1 monoclonal antibody to tumour necrosis factor that is administered subcutaneously and is indicated for use in adults with CD who have had an inadequate response to conventional therapy. However,clinical trial experience with adalimumab in CD is limited to adults" [1]. "It was also found as an adverse reaction in one patient on a controlled experiment to evaluate the efficacy of adalimumab in juvenile idiopathic arthritis-associated uveitis" [3].

references:
1. http://journals.lww.com/jpgn/Fulltext/2008/02000/Adalimumab_Induces_and_Maintains_Remission_in.14.aspx
2. http://www.mayoclinic.org/diseases-conditions/crohns-disease/basics/definition/con-20032061
3. http://rheumatology.oxfordjournals.org/content/47/3/339.full

### 19) histamine phosphate - hypotension (proximity: 0.1875) - ADR

"**Side effects** can lead to hypertension, **hypotension**, headache, dizziness, nervousness and tachycardia. Large overdoses can lead to seizures" [1].

references:
1. DrugBank

### 20) conjugated estrogens - dyspepsia (proximity: 0.2105) - ADR

Indications: "For the treatment of moderate to severe vasomotor symptoms associated with the menopause [...]"[1]

"The most common adverse drug events (with an incidence of at least 5%) associated with **CE/BZA** have included muscle spasms, dizziness, nausea, diarrhea, **dyspepsia**, upper abdominal pain, neck pain, and throat pain." [2]

references:
1. DrugBank
2. http://www.ncbi.nlm.nih.gov/pmc/articles/PMC4357350/

### 21) donepezil - rhinitis (proximity: 0.2) - ADR

Indications: "For the palliative treatment of mild to moderate dementia of the Alzheimer's type."[1]
"The most common adverse events included nausea, diarrhoea, headache, insomnia, dizziness, **rhinitis**, vomiting, asthenia/fatigue and anorexia."[2]

references:
1. DrugBank
2. http://www.ncbi.nlm.nih.gov/pubmed/12469988

### 22) scopolamine - neurotoxicity (proximity: 0.2) - Possible ADR

Indications: "For the treatment of excessive salivation, colicky abdominal pain, bradycardia, sialorrhoea, diverticulitis, irritable bowel syndrome and motion sickness."[1]
Sopolamine is known and used for inducing amnesia and dementia in lab animals(e.g. rats and monkeys), though not clear it causes the same in humans [2].

references:
1. DrugBank
2. http://jop.sagepub.com/content/6/3/382.abstract

### 23) ribavirin - myopathy (proximity: 0.2) - Possible ADR (unknown, 1 case)

Indications: "For the treatment of chronic hepatitis C and for respiratory syncytial virus (RSV)."[1]
"Adverse events induced by interferon therapy are numerous but myopathy is rare and has not been described with the use of pegylated interferon-a. We report the case of a 33-year-old Caucasian man who was successfully treated for acute hepatitis C with the combination of pegylated interferon-a2b and **ribavirin**, and who during treatment developed **myopathy** which proved reversible."[2]

references:
1. DrugBank
2. http://onlinelibrary.wiley.com/store/10.1046/j.1365-2893.2003.00478.x/asset/j.1365-2893.2003.00478.x.pdf?v=1&t=id69jh7g&s=1a95bda8c2380a01e0e8fe0b9ce4d1ea40619beb

### 24) tiotropium - myopathy (proximity: 0.2) - Nothing

Indications: "Used in the management of chronic obstructive pulmonary disease (COPD)."[1]
references:
1. DrugBank

### 25) vardenafil - hypotension (proximity: 0.1875) - ADR

Indications: "Used for the treatment of erectile dysfunction"[1]
Like sildenafil, **vardenafil** has a slightly **hypotensive effect**, maximal 5-10 mmHg average[2]

references:
1. DrugBank
2. http://www.ncbi.nlm.nih.gov/pubmed/12435622

# Preliminary evidence for drug-symptoms associations in indirect associations (Figure 6)

### 1) cinnarizine - hypotension (metric closure proximity: 0.21) - Nothing
Indications: "For the treatment of vertigo/meniere's disease, nausea and vomiting, motion sickness and also useful for vestibular symptoms of other origins."[1]

references:
1. DrugBank

### 2) bethanechol - hypotension (metric closure proximity: 0.21) - ADR
Indications: "For the treatment of acute postoperative and postpartum nonobstructive (functional) urinary retention and for neurogenic atony of the urinary bladder with retention. It may cause **hypotension**, cardiac rate changes, and bronchial spasms."[1]

references:
1. DrugBank

### 3) cabergoline - hypotension (metric closure proximity: 0.21) - Possible ADR
Indications: "For the treatment of hyperprolactinemic disorders, either idiopathic or due to prolactinoma (prolactin-secreting adenomas). May also be used to manage symptoms of Parkinsonian Syndrome as monotherapy during initial symptomatic management or as an adjunct to levodopa therapy during advanced stages of disease."[1]
"Initial doses higher than 1.0 mg may produce **orthostatic hypotension**. Care should be exercised when administering Cabergoline with other medications known to lower blood pressure."[2]

references:
2. DrugBank
3. http://www.drugs.com/pro/cabergoline.html

### 4) primidone - hypotension (metric closure proximity: 0.196) - ADR (overdose)
Indications: "For the treatment of epilepsy"[1]
Case: Laboratory studies demonstrated extremely high serum concentrations of methsuximide (98.5 mg/liter) and **primidone** (62 mg/liter), and she was admitted to overdose of primidone. [...] she developed **hypotension** with urine output decreased to 3ml/h and a systolic blood pressure of 50-60 mmHg [2]. Other cases, see [3]

references:
1. DrugBank
2. http://www.clinchem.org/content/22/6/915.full.pdf
3. http://toxnet.nlm.nih.gov/cgi-bin/sis/search/a?dbs+hsdb:@term+@DOCNO+3169

### 5) desmopressin - hypotension (metric closure proximity: 0.196) - ADR

Indications: "Oral formulations may be used to manage primary nocturnal enuresis in adults and vasopressin sensitive diabetes insipidus, and for control of temporary polyuria and polydipsia following head trauma or surgery in the pituitary region. Intranasal and parenteral formulations may be used to manage spontaneous or trauma-induced bleeds (e.g. hemarthrosis, intramuscular hematoma, mucosal bleeding) in patients with hemophilia A or von Willebrand's disease Type I. May also be used parenterally to prevent or treat bleeding in patients with uremia."[1]

"**Desmopressin acetate** is used to reduce blood loss after cardiac surgery. However, there have been reports that **hypotension** can occur with infusion of desmopressin and that postoperative blood loss is not reduced. [Results shows] a 20% or greater decrease in mean arterial pressure was observed in 7 of 20 patients receiving desmopressin."

references:
1. DrugBank
2. http://europepmc.org/abstract/med/2042789

### 6) sevelamer - hypotension (metric closure proximity: 0.196) - Possible ADR (case)

Indications: "For the control of serum phosphorus in patients with Chronic Kidney Disease (CKD) on hemodialysis."[1]

"A 62-year-old woman presented with bleeding per rectum for one day. She reported a history of constipation which occurred 1 mo ago when she started taking Sevelamer.[...] Besides Sevelamer, she was also taking Clopidogrel but never had similar episodes of bleeding before. Physical examination revealed tachycardia and **hypotension**."

references:
1. DrugBank
2. http://www.ncbi.nlm.nih.gov/pmc/articles/PMC2708383/

### 7) atomoxetine - hypotension (metric closure proximity: 0.188) - Possible ADR (case)

Indications: "For the treatment of Attention-Deficit/Hyperactivity Disorder (ADHD) alone or in combination with behavioral treatment, as an adjunct to psychological, educational, social, and other remedial measures."[1]

Case: "The common side effects reported with the use of atomoxetine include mainly GI disturbances. **Cardiovascular side effects are less commonly reported**. The increase in the noradrenergic tone may explain some of the side effects noted with the use of this medication. Here, we present a case of a patient who presented with syncope, **orthostatic hypotension**, and tachycardia and discuss the various clinical implications based on the pharmacokinetics and pharmacodynamics of the drug."

references:
1. DrugBank
2. http://www.hindawi.com/journals/crim/2011/952584/abs/

### 8) vasopressin - hypotension (metric closure proximity: 0.184) - Unknown

Indications: "For the treatment of enuresis, polyuria, diabetes insipidus, polydipsia and oesophageal varices with bleeding"[1]

references:
1. DrugBank

### 9) montelukast - angina (metric closure proximity: 0.180) - Symptom

"Montelukast is a leukotriene receptor antagonist (LTRA) used for the maintenance treatment of asthma and to relieve symptoms of seasonal allergies" [1] and are also "**useful in treating angina**, cerebral spasm, glomerular nephritis, hepatitis, endotoxemia, uveitis, and allograft rejection".[2]

references:
1. DrugBank
2. Patent: https://www.google.com/patents/US5565473

### 10) lidocaine - angina (metric closure proximity: 0.177) - Symptom

Indications: "Indications: "For production of local or regional anesthesia by infiltration techniques such as percutaneous injection and intravenous regional anesthesia by peripheral nerve block techniques such as brachial plexus and intercostal and by central neural techniques such as lumbar and caudal epidural blocks."[1]

"**Lidocaine** is an anesthetic agent indicated for production of local or regional anesthesia and in the **treatment of ventricular tachycardia** occurring during cardiac manipulation, such as surgery or catheterization, or which may occur during acute myocardial infarction, digitalis toxicity, or other cardiac diseases."[1]

references:
1. DrugBank

### 11) liraglutide - hypotension (metric closure proximity: 0.174) - ADR

Indications: "For use in/treatment of diabetes mellitus type 2."[1]
Cardiovascular Side Effects: Very common (10% or more): Increases in mean resting heart rate; Common (1% to 10%): **Hypotension**[2].

references:
1. DrugBank
2. http://www.drugs.com/sfx/liraglutide-side-effects.html

### 12) naloxone - rhinitis (metric closure proximity: 0.167) - ADR

Indications: "For the complete or partial reversal of narcotic depression, including respiratory depression, induced by opioids including natural and synthetic narcotics, propoxyphene, methadone and the narcotic-antagonist analgesics: nalbuphine, pentazocine and butorphanol. It is also indicated for the

diagnosis of suspected acute opioid overdose. It may also be used as an adjunctive agent to increase blood pressure in the management of septic shock."[1]

"**Naloxone** may precipitate withdrawal in patients receiving opioids. Withdrawal is characterized by nausea, vomiting, sweating, lacrimation, **rhinorrhea**, cramping[...]"[2]

references:
1. DrugBank
2. http://www.drugs.com/naloxone.html

### 13) cinnarizine - angina (metric closure proximity: 0.163) - Unknown

Indications: "For the treatment of vertigo/meniere's disease, nausea and vomiting, motion sickness and also useful for vestibular symptoms of other origins."[1]

references:
1. DrugBank

### 14) bethanechol - angina (metric closure proximity: 0.163) - Unknown

Indications: "For the treatment of acute postoperative and postpartum nonobstructive (functional) urinary retention and for neurogenic atony of the urinary bladder with retention."[1]

references:
1. DrugBank

### 15) cabergoline - angina (metric closure proximity: 0.163) - Unknown

Indications: "For the treatment of hyperprolactinemic disorders, either idiopathic or due to prolactinoma (prolactin-secreting adenomas). May also be used to manage symptoms of Parkinsonian Syndrome as monotherapy during initial symptomatic management or as an adjunct to levodopa therapy during advanced stages of disease."[1]

references:
1. DrugBank

### 16) allopurinol - myopathy (metric closure proximity: 0.158) - Unknown

Indications: "For the treatment of hyperuricemia associated with primary or secondary gout. Also indicated for the treatment of primary or secondary uric acid nephropathy, with or without the symptoms of gout, as well as chemotherapy-induced hyperuricemia and recurrent renal calculi."[1]

references:
1. DrugBank

### 17) cilostazol - hypotension (metric closure proximity: 0.156) - ADR (overdose)

Indications: "For the reduction of symptoms of intermittent claudication (pain in the legs that occurs with walking and disappears with rest)."[1]

"Information on acute overdosage with cilostazol in humans is limited. The signs and symptoms of an **acute overdose** can be anticipated to be those of excessive pharmacologic effect: severe headache, diarrhea, **hypotension**, tachycardia, and possibly cardiac arrhythmias."[1]

references:
1. DrugBank

### 18) follitropin beta - hypotension (metric closure proximity: 0.156) - Unknown

Indications: "For treatment of female infertility"[1]

references:
4. DrugBank

### 19) tetrabenazine - hypotension (metric closure proximity: 0.156) - Unknown

Indications: "Treatment of hyperkinetic movement disorders like chorea in Huntington's disease, hemiballismus, senile chorea, Tourette syndrome and other tic disorders, and tardive dyskinesia"[1]

references:
5. DrugBank

### 20) nevirapine - hypotension (metric closure proximity: 0.156) - Unknown

Indications: "For use in combination with other antiretroviral drugs in the ongoing treatment of HIV-1 infection."[1]

references:
6. DrugBank

### 21) flumazenil - hypotension (metric closure proximity: 0.156) - Unknown

Indications: "For the complete or partial reversal of the sedative effects of benzodiazepines in cases where general anesthesia has been induced and/or maintained with benzodiazepines, and where sedation has been produced with benzodiazepines for diagnostic and therapeutic procedures. Also for the management of benzodiazepine overdose as an adjunct for appropriate supportive and symptomatic measures."[1]

references:
7. DrugBank

### 22) alendronate - hypotension (metric closure proximity: 0.156) - Unknown

Indications: "For the treatment and prevention of osteoporosis in women and Paget's disease of bone in both men and women."[1]

references:
8. DrugBank

### 23) betahistine - hypotension (metric closure proximity: 0.156) - Unknown

Indications: "For the reduction of episodes of vertigo association with Ménière's disease."[1]

references:
9. DrugBank

### 24) natamycin - hypotension (metric closure proximity: 0.156) - Unknown

Indications: "For the treatment of fungal blepharitis, conjunctivitis, and keratitis caused by susceptible organisms including Fusarium solani keratitis."[1]

references:
10. DrugBank

### 25) prasugrel - hypotension (metric closure proximity: 0.156) - Unknown

Indications: "ndicated in combination with acetylsalicylic acid (ASA) to prevent atherothrombotic events in patients with acute coronary syndrome (ACS) who are to be managed with percutaneous coronary intervention (PCI)"[1]

references:
11. DrugBank

# Psoriasis network identified with PC 4 and colored by PC5

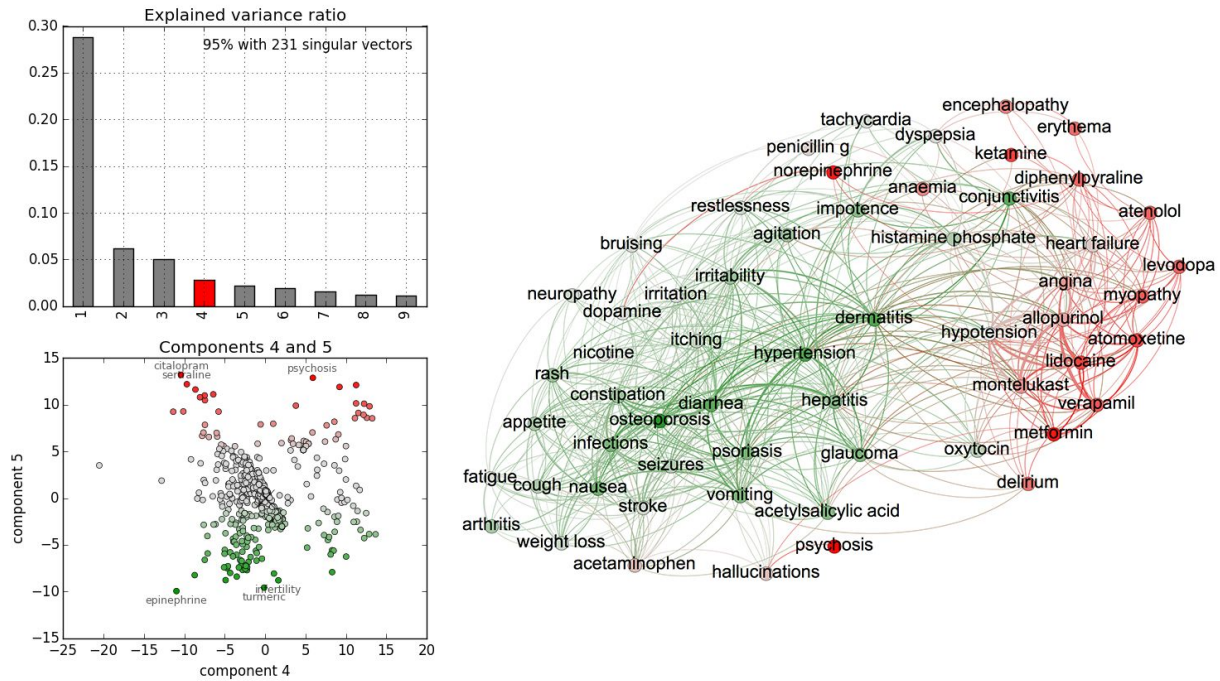

PC5 further characterizes the Psoriasis network identified with PC 4, uncovering two subnetwork clusters. Red (green) nodes are (anti-)correlated with PC5. Red nodes are associated with a number of drugs (such as metformin for Diabetes) as well as is related to hypotension, angina, and heart failure. The green nodes are, in turn, associated hypertension as well as the cluster of conditions linked to psoriasis such as glaucoma, hepatitis, arthritis, seizures and stroke. Additionally, red nodes are associated with depression drugs citalopram and sertraline which are external to subnetwork of PC4 but strongly correlated with PC5, as well as psychosis and related conditions. Green nodes are also associated with many NP terms which are external to subnetwork of PC4 but strongly anti-correlated with PC5.

Some of the NP anti-correlated with PC5 are: turmeric, aloe, acai, flaxseed, elderberry, echinacea, peppermint oil, licorice, chamomile, fenugreek, valerian.

# Queries on the 1 Week Distance and Transitive Closure Network

Query 1W: **'pain'** and **'cannabis'**
Ordering: **avg**

Proximity Network Results

|           | pain  | cannabis | (max) | (min) | (avg) |
|-----------|-------|----------|-------|-------|-------|
| pain      | 1.000 | 0.058    | 1.000 | 0.058 | 0.529 |
| cannabis  | 0.058 | 1.000    | 1.000 | 0.058 | 0.529 |
| anxiety   | 0.160 | 0.034    | 0.160 | 0.034 | 0.097 |
| insomnia  | 0.170 | 0.019    | 0.170 | 0.019 | 0.095 |
| suicide   | 0.156 | 0.025    | 0.156 | 0.025 | 0.090 |
| mood      | 0.121 | 0.032    | 0.121 | 0.032 | 0.076 |
| suicidal  | 0.130 | 0.012    | 0.130 | 0.012 | 0.071 |
| fluoxetine| 0.099 | 0.028    | 0.099 | 0.028 | 0.063 |
| ginger    | 0.083 | 0.025    | 0.083 | 0.025 | 0.054 |
| headache  | 0.086 | 0.013    | 0.086 | 0.013 | 0.050 |
| cinnamon  | 0.079 | 0.020    | 0.079 | 0.020 | 0.049 |
| garlic    | 0.068 | 0.023    | 0.068 | 0.023 | 0.046 |
| flu       | 0.067 | 0.018    | 0.067 | 0.018 | 0.042 |
| migraine  | 0.075 | 0.008    | 0.075 | 0.008 | 0.042 |
| burning   | 0.067 | 0.015    | 0.067 | 0.015 | 0.041 |
| arthritis | 0.069 | 0.007    | 0.069 | 0.007 | 0.038 |
| fatigue   | 0.065 | 0.008    | 0.065 | 0.008 | 0.036 |
| soy       | 0.056 | 0.015    | 0.056 | 0.015 | 0.035 |
| caffeine  | 0.054 | 0.012    | 0.054 | 0.012 | 0.033 |
| lavender  | 0.050 | 0.012    | 0.050 | 0.012 | 0.031 |

Transitive Closure (proximity) Networks Results

|           | pain  | cannabis | (max) | (min) | (avg) |
|-----------|-------|----------|-------|-------|-------|
| pain      | 1.000 | 0.058    | 1.000 | 0.058 | 0.529 |
| cannabis  | 0.058 | 1.000    | 1.000 | 0.058 | 0.529 |
| insomnia  | 0.170 | 0.045    | 0.170 | 0.045 | 0.108 |
| anxiety   | 0.160 | 0.045    | 0.160 | 0.045 | 0.102 |
| suicide   | 0.156 | 0.044    | 0.156 | 0.044 | 0.100 |
| suicidal  | 0.132 | 0.042    | 0.132 | 0.042 | 0.087 |
| mood      | 0.121 | 0.041    | 0.121 | 0.041 | 0.081 |
| anorexia  | 0.103 | 0.039    | 0.103 | 0.039 | 0.071 |
| fluoxetine| 0.099 | 0.038    | 0.099 | 0.038 | 0.068 |
| headache  | 0.086 | 0.036    | 0.086 | 0.036 | 0.061 |

---

Query 1W: **'fluoxetine'** and **'anorexia'**
Ordering: **min**

Proximity Network Results

|          | fluoxetine | anorexia | (max) | (min) | (avg) |
|----------|------------|----------|-------|-------|-------|
| suicidal | 0.075      | 0.111    | 0.111 | 0.075 | 0.093 |

```
suicide          0.073    0.093 0.093 0.073 0.083
anxiety          0.071    0.225 0.225 0.071 0.148
pain             0.099    0.041 0.099 0.041 0.070
mood             0.089    0.028 0.089 0.028 0.058
cinnamon         0.066    0.022 0.066 0.022 0.044
fluoxetine       1.000    0.020 1.000 0.020 0.510
anorexia         0.020    1.000 1.000 0.020 0.510
insomnia         0.081    0.019 0.081 0.019 0.050
soy              0.064    0.014 0.064 0.014 0.039
headache         0.075    0.014 0.075 0.014 0.044
mania            0.022    0.011 0.022 0.011 0.016
chia             0.034    0.010 0.034 0.010 0.022
cannabis         0.028    0.009 0.028 0.009 0.018
ginger           0.070    0.008 0.070 0.008 0.039
psychosis        0.017    0.008 0.017 0.008 0.012
green tea        0.039    0.008 0.039 0.008 0.023
burning          0.075    0.007 0.075 0.007 0.041
appetite         0.037    0.006 0.037 0.006 0.022
garlic           0.062    0.006 0.062 0.006 0.034


Transitive Closure (proximity) Networks Results
           fluoxetine  anorexia  (max)  (min)  (avg)
pain            0.099    0.103 0.103 0.099 0.101
insomnia        0.081    0.079 0.081 0.079 0.080
suicidal        0.075    0.114 0.114 0.075 0.094
suicide         0.073    0.116 0.116 0.073 0.094
anxiety         0.071    0.225 0.225 0.071 0.148
mood            0.089    0.059 0.089 0.059 0.074
fluoxetine      1.000    0.057 1.000 0.057 0.529
anorexia        0.057    1.000 1.000 0.057 0.529
migraine        0.054    0.050 0.054 0.050 0.052
headache        0.075    0.049 0.075 0.049 0.062
```

---

Query 1W: **'mood', 'cannabis', 'fluoxetine'**
Ordering: **avg**

```
Proximity Network Results
           cannabis  mood  fluoxetine  (max)  (min)  (avg)
mood          0.032  1.000      0.089  1.000  0.032  0.373
fluoxetine    0.028  0.089      1.000  1.000  0.028  0.372
cannabis      1.000  0.032      0.028  1.000  0.028  0.353
pain          0.058  0.121      0.099  0.121  0.058  0.092
insomnia      0.019  0.083      0.081  0.083  0.019  0.061
anxiety       0.034  0.071      0.071  0.071  0.034  0.059
headache      0.013  0.086      0.075  0.086  0.013  0.058
ginger        0.025  0.074      0.070  0.074  0.025  0.056
cinnamon      0.020  0.082      0.066  0.082  0.020  0.056
suicide       0.025  0.064      0.073  0.073  0.025  0.054
```

```
burning          0.015  0.066         0.075  0.075  0.015  0.052
garlic           0.023  0.064         0.062  0.064  0.023  0.050
suicidal         0.012  0.055         0.075  0.075  0.012  0.047
flu              0.018  0.062         0.058  0.062  0.018  0.046
caffeine         0.012  0.063         0.060  0.063  0.012  0.045
soy              0.015  0.054         0.064  0.064  0.015  0.044
fever            0.009  0.045         0.053  0.053  0.009  0.036
cough            0.008  0.041         0.041  0.041  0.008  0.030
lavender         0.012  0.039         0.035  0.039  0.012  0.029
migraine         0.008  0.046         0.030  0.046  0.008  0.028

Transitive Closure (proximity) Networks Results
            cannabis  mood  fluoxetine  (max)  (min)  (avg)
mood           0.041  1.000       0.089  1.000  0.041  0.377
fluoxetine     0.038  0.089       1.000  1.000  0.038  0.376
cannabis       1.000  0.041       0.038  1.000  0.038  0.360
pain           0.058  0.121       0.099  0.121  0.058  0.092
insomnia       0.045  0.083       0.081  0.083  0.045  0.070
headache       0.036  0.086       0.075  0.086  0.036  0.066
suicide        0.044  0.073       0.073  0.073  0.044  0.063
anxiety        0.045  0.074       0.071  0.074  0.045  0.063
suicidal       0.042  0.067       0.075  0.075  0.042  0.061
cinnamon       0.035  0.082       0.066  0.082  0.035  0.061
```

Query 1W: **'hepatitis'**
Ordering: **avg**

Proximity Network Results

```
                     hepatitis  (max)  (min)  (avg)
hepatitis               1.000   1.000  1.000  1.000
glaucoma                0.195   0.195  0.195  0.195
hypertension            0.171   0.171  0.171  0.171
diarrhea                0.169   0.169  0.169  0.169
impotence               0.169   0.169  0.169  0.169
seizures                0.163   0.163  0.163  0.163
dermatitis              0.155   0.155  0.155  0.155
irritability            0.145   0.145  0.145  0.145
psoriasis               0.144   0.144  0.144  0.144
hypotension             0.130   0.130  0.130  0.130
vomiting                0.126   0.126  0.126  0.126
constipation            0.124   0.124  0.124  0.124
stroke                  0.122   0.122  0.122  0.122
infections              0.122   0.122  0.122  0.122
osteoporosis            0.114   0.114  0.114  0.114
acetylsalicylic acid    0.113   0.113  0.113  0.113
infertility             0.113   0.113  0.113  0.113
histamine phosphate     0.107   0.107  0.107  0.107
itching                 0.105   0.105  0.105  0.105
stinging                0.103   0.103  0.103  0.103
```

```
Transitive Closure (proximity) Networks Results
              hepatitis  (max)  (min)  (avg)
hepatitis       1.000    1.000  1.000  1.000
glaucoma        0.195    0.195  0.195  0.195
hypertension    0.171    0.171  0.171  0.171
diarrhea        0.169    0.169  0.169  0.169
impotence       0.169    0.169  0.169  0.169
seizures        0.163    0.163  0.163  0.163
dermatitis      0.155    0.155  0.155  0.155
irritability    0.145    0.145  0.145  0.145
psoriasis       0.144    0.144  0.144  0.144
hypotension     0.130    0.130  0.130  0.130
```

---

Query 1W:`psoriasis`,`heart failure`
Ordering: **min**

```
Proximity Network Results
                      psoriasis  heart failure  (max)  (min)  (avg)
dermatitis               0.125       0.115      0.125  0.115  0.120
montelukast              0.083       0.114      0.114  0.083  0.099
hypertension             0.159       0.068      0.159  0.068  0.113
oxytocin                 0.060       0.066      0.066  0.060  0.063
hypotension              0.058       0.100      0.100  0.058  0.079
hepatitis                0.144       0.057      0.144  0.057  0.101
conjugated estrogens     0.051       0.065      0.065  0.051  0.058
lidocaine                0.050       0.091      0.091  0.050  0.071
conjunctivitis           0.049       0.079      0.079  0.049  0.064
drowsiness               0.048       0.050      0.050  0.048  0.049
irritability             0.112       0.048      0.112  0.048  0.080
itching                  0.109       0.046      0.109  0.046  0.077
allopurinol              0.046       0.160      0.160  0.046  0.103
heart failure            0.043       1.000      1.000  0.043  0.522
psoriasis                1.000       0.043      1.000  0.043  0.522
bruising                 0.050       0.043      0.050  0.043  0.047
histamine phosphate      0.043       0.083      0.083  0.043  0.063
tachycardia              0.043       0.053      0.053  0.043  0.048
verapamil                0.042       0.077      0.077  0.042  0.060
hydrocortisone           0.050       0.041      0.050  0.041  0.046

Transitive Closure (proximity) Networks Results
                   psoriasis  heart failure  (max)  (min)  (avg)
dermatitis           0.125       0.115       0.125  0.115  0.120
montelukast          0.083       0.125       0.125  0.083  0.104
dry skin             0.087       0.082       0.087  0.082  0.085
hypotension          0.081       0.130       0.130  0.081  0.106
elderberry           0.081       0.077       0.081  0.077  0.079
irritability         0.112       0.077       0.112  0.077  0.094
hypertension         0.159       0.074       0.159  0.074  0.116
```

```
naloxone              0.073          0.145 0.145 0.073 0.109
allopurinol           0.072          0.160 0.160 0.072 0.116
peppermint oil        0.075          0.072 0.075 0.072 0.074
```

---

Query 1W:`**psoriasis**`,`**hepatitis**`  
Ordering: **min**

```
Proximity Network Results
                    psoriasis  hepatitis  (max)  (min)  (avg)
diarrhea              0.160      0.169    0.169  0.160  0.165
hypertension          0.159      0.171    0.171  0.159  0.165
psoriasis             1.000      0.144    1.000  0.144  0.572
hepatitis             0.144      1.000    1.000  0.144  0.572
glaucoma              0.135      0.195    0.195  0.135  0.165
dermatitis            0.125      0.155    0.155  0.125  0.140
constipation          0.145      0.124    0.145  0.124  0.135
stroke                0.140      0.122    0.140  0.122  0.131
infections            0.183      0.122    0.183  0.122  0.152
irritability          0.112      0.145    0.145  0.112  0.128
seizures              0.108      0.163    0.163  0.108  0.135
itching               0.109      0.105    0.109  0.105  0.107
osteoporosis          0.102      0.114    0.114  0.102  0.108
irritation            0.105      0.100    0.105  0.100  0.103
infertility           0.094      0.113    0.113  0.094  0.104
vomiting              0.091      0.126    0.126  0.091  0.108
allergic              0.088      0.084    0.088  0.084  0.086
montelukast           0.083      0.094    0.094  0.083  0.088
nausea                0.076      0.080    0.080  0.076  0.078
acetylsalicylic acid  0.073      0.113    0.113  0.073  0.093

Transitive Closure (proximity) Networks Results
            psoriasis  hepatitis  (max)  (min)  (avg)
diarrhea      0.160      0.169    0.169  0.160  0.165
hypertension  0.159      0.171    0.171  0.159  0.165
psoriasis     1.000      0.144    1.000  0.144  0.572
hepatitis     0.144      1.000    1.000  0.144  0.572
glaucoma      0.135      0.195    0.195  0.135  0.165
dermatitis    0.125      0.155    0.155  0.125  0.140
constipation  0.145      0.124    0.145  0.124  0.135
stroke        0.140      0.122    0.140  0.122  0.131
infections    0.183      0.122    0.183  0.122  0.152
irritability  0.112      0.145    0.145  0.112  0.128
```

---

Query 1W:`**psoriasis**`,`**seizures**`  
Ordering: **min**

```
Proximity Network Results
              psoriasis  seizures  (max)  (min)  (avg)
infections       0.183    0.205    0.205  0.183  0.194
hepatitis        0.144    0.163    0.163  0.144  0.153
stroke           0.140    0.189    0.189  0.140  0.164
diarrhea         0.160    0.122    0.160  0.122  0.141
itching          0.109    0.109    0.109  0.109  0.109
seizures         0.108    1.000    1.000  0.108  0.554
psoriasis        1.000    0.108    1.000  0.108  0.554
constipation     0.145    0.102    0.145  0.102  0.124
osteoporosis     0.102    0.104    0.104  0.102  0.103
glaucoma         0.135    0.101    0.135  0.101  0.118
irritability     0.112    0.099    0.112  0.099  0.105
infertility      0.094    0.150    0.150  0.094  0.122
vomiting         0.091    0.150    0.150  0.091  0.120
irritation       0.105    0.091    0.105  0.091  0.098
hypertension     0.159    0.087    0.159  0.087  0.123
dermatitis       0.125    0.084    0.125  0.084  0.104
allergic         0.088    0.083    0.088  0.083  0.086
acne             0.119    0.081    0.119  0.081  0.100
ibuprofen        0.076    0.076    0.076  0.076  0.076
appetite         0.076    0.078    0.078  0.076  0.077


Transitive Closure (proximity) Networks Results
              psoriasis  seizures  (max)  (min)  (avg)
infections       0.183    0.205    0.205  0.183  0.194
hepatitis        0.144    0.163    0.163  0.144  0.153
stroke           0.140    0.189    0.189  0.140  0.164
diarrhea         0.160    0.122    0.160  0.122  0.141
constipation     0.145    0.121    0.145  0.121  0.133
acne             0.119    0.113    0.119  0.113  0.116
itching          0.109    0.109    0.109  0.109  0.109
psoriasis        1.000    0.108    1.000  0.108  0.554
seizures         0.108    1.000    1.000  0.108  0.554
osteoporosis     0.102    0.104    0.104  0.102  0.103
```

---

Query 1W:`**psoriasis**`,`**heart failure**`,`**stroke**`
Ordering: **min**

```
Proximity Network Results
                psoriasis  stroke  heart failure  (max)  (min)  (avg)
hypertension      0.159    0.107       0.068      0.159  0.068  0.111
dermatitis        0.125    0.065       0.115      0.125  0.065  0.102
hepatitis         0.144    0.122       0.057      0.144  0.057  0.108
irritability      0.112    0.068       0.048      0.112  0.048  0.076
itching           0.109    0.080       0.046      0.109  0.046  0.078
psoriasis         1.000    0.140       0.043      1.000  0.043  0.394
bruising          0.050    0.052       0.043      0.052  0.043  0.048
```

```
montelukast             0.083  0.043         0.114  0.114  0.043  0.080
oxytocin                0.060  0.042         0.066  0.066  0.042  0.056
glaucoma                0.135  0.109         0.040  0.135  0.040  0.094
drowsiness              0.048  0.039         0.050  0.050  0.039  0.046
hydrocortisone          0.050  0.039         0.041  0.050  0.039  0.043
diarrhea                0.160  0.155         0.037  0.160  0.037  0.118
vomiting                0.091  0.131         0.035  0.131  0.035  0.086
dizziness               0.042  0.074         0.035  0.074  0.035  0.050
lethargy                0.035  0.045         0.043  0.045  0.035  0.041
hypotension             0.058  0.034         0.100  0.100  0.034  0.064
agitation               0.040  0.033         0.041  0.041  0.033  0.038
acetylsalicylic acid    0.073  0.064         0.032  0.073  0.032  0.056
seizures                0.108  0.189         0.030  0.189  0.030  0.109

Transitive Closure (proximity) Networks Results
              psoriasis  stroke  heart failure  (max)  (min)  (avg)
irritability    0.112    0.086         0.077   0.112  0.077  0.091
dermatitis      0.125    0.077         0.115   0.125  0.077  0.106
hypertension    0.159    0.107         0.074   0.159  0.074  0.113
dry skin        0.087    0.073         0.082   0.087  0.073  0.081
itching         0.109    0.080         0.071   0.109  0.071  0.087
hepatitis       0.144    0.122         0.071   0.144  0.071  0.112
irritation      0.105    0.077         0.068   0.105  0.068  0.083
hypotension     0.081    0.067         0.130   0.130  0.067  0.093
elderberry      0.081    0.067         0.077   0.081  0.067  0.075
montelukast     0.083    0.066         0.125   0.125  0.066  0.091
```

---

Query 1W:`**psoriasis**',`**heart failure**',`**stroke**'
Ordering: **avg**

```
Proximity Network Results
              psoriasis  stroke  heart failure  (max)  (min)  (avg)
psoriasis       1.000    0.140         0.043   1.000  0.043  0.394
stroke          0.140    1.000         0.027   1.000  0.027  0.389
heart failure   0.043    0.027         1.000   1.000  0.027  0.357
infections      0.183    0.213         0.024   0.213  0.024  0.140
diarrhea        0.160    0.155         0.037   0.160  0.037  0.118
hypertension    0.159    0.107         0.068   0.159  0.068  0.111
seizures        0.108    0.189         0.030   0.189  0.030  0.109
hepatitis       0.144    0.122         0.057   0.144  0.057  0.108
constipation    0.145    0.141         0.020   0.145  0.020  0.102
dermatitis      0.125    0.065         0.115   0.125  0.065  0.102
glaucoma        0.135    0.109         0.040   0.135  0.040  0.094
vomiting        0.091    0.131         0.035   0.131  0.035  0.086
montelukast     0.083    0.043         0.114   0.114  0.043  0.080
itching         0.109    0.080         0.046   0.109  0.046  0.078
angina          0.031    0.021         0.179   0.179  0.021  0.077
acne            0.119    0.099         0.010   0.119  0.010  0.076
```

```
irritability          0.112   0.068           0.048   0.112   0.048   0.076
allopurinol           0.046   0.022           0.160   0.160   0.022   0.076
infertility           0.094   0.121           0.008   0.121   0.008   0.074
nausea                0.076   0.123           0.016   0.123   0.016   0.071


Transitive Closure (proximity) Networks Results
                 psoriasis   stroke   heart failure   (max)   (min)   (avg)
psoriasis           1.000    0.140           0.064    1.000   0.064   0.401
stroke              0.140    1.000           0.048    1.000   0.048   0.396
heart failure       0.064    0.048           1.000    1.000   0.048   0.371
infections          0.183    0.213           0.052    0.213   0.052   0.149
diarrhea            0.160    0.155           0.066    0.160   0.066   0.127
seizures            0.108    0.189           0.052    0.189   0.052   0.116
constipation        0.145    0.141           0.060    0.145   0.060   0.115
hypertension        0.159    0.107           0.074    0.159   0.074   0.113
hepatitis           0.144    0.122           0.071    0.144   0.071   0.112
dermatitis          0.125    0.077           0.115    0.125   0.077   0.106
```

---

Query 1W:`**psoriasis**`,`**metformin**`
Ordering: **min**

```
Proximity Network Results
                    psoriasis   metformin   (max)   (min)   (avg)
montelukast             0.083       0.098   0.098   0.083   0.091
hypertension            0.159       0.070   0.159   0.070   0.114
dermatitis              0.125       0.062   0.125   0.062   0.093
hypotension             0.058       0.109   0.109   0.058   0.084
hepatitis               0.144       0.053   0.144   0.053   0.098
oxytocin                0.060       0.050   0.060   0.050   0.055
lidocaine               0.050       0.143   0.143   0.050   0.097
vomiting                0.091       0.049   0.091   0.049   0.070
glaucoma                0.135       0.047   0.135   0.047   0.091
allopurinol             0.046       0.118   0.118   0.046   0.082
heart failure           0.043       0.048   0.048   0.043   0.045
verapamil               0.042       0.111   0.111   0.042   0.077
diarrhea                0.160       0.042   0.160   0.042   0.101
ibuprofen               0.076       0.042   0.076   0.042   0.059
naloxone                0.040       0.128   0.128   0.040   0.084
agitation               0.040       0.040   0.040   0.040   0.040
acetylsalicylic acid    0.073       0.038   0.073   0.038   0.056
nicotine                0.038       0.033   0.038   0.033   0.036
angina                  0.031       0.070   0.070   0.031   0.051
metformin               0.031       1.000   1.000   0.031   0.516


Transitive Closure (proximity) Networks Results
                    psoriasis   metformin   (max)   (min)   (avg)
montelukast             0.083       0.098   0.098   0.083   0.091
```

| | | | | | |
|---|---|---|---|---|---|
| hypotension | 0.081 | | 0.109 | 0.109 | 0.081 | 0.095 |
| naloxone | 0.073 | | 0.128 | 0.128 | 0.073 | 0.100 |
| allopurinol | 0.072 | | 0.118 | 0.118 | 0.072 | 0.095 |
| hypertension | 0.159 | | 0.070 | 0.159 | 0.070 | 0.114 |
| histamine phosphate | 0.069 | | 0.078 | 0.078 | 0.069 | 0.074 |
| dermatitis | 0.125 | | 0.068 | 0.125 | 0.068 | 0.097 |
| ribavirin | 0.067 | | 0.100 | 0.100 | 0.067 | 0.084 |
| tiotropium | 0.067 | | 0.100 | 0.100 | 0.067 | 0.084 |
| angina | 0.066 | | 0.092 | 0.092 | 0.066 | 0.079 |

---

Query 1W:`**psoriasis**`,`**insulin glargine**`
Ordering: **min**

Proximity Network Results

| | psoriasis | insulin glargine | (max) | (min) | (avg) |
|---|---|---|---|---|---|
| lethargy | 0.035 | 0.016 | 0.035 | 0.016 | 0.025 |
| metformin | 0.031 | 0.013 | 0.031 | 0.013 | 0.022 |
| tenderness | 0.019 | 0.010 | 0.019 | 0.010 | 0.014 |
| vomiting | 0.091 | 0.009 | 0.091 | 0.009 | 0.050 |
| weakness | 0.036 | 0.009 | 0.036 | 0.009 | 0.022 |
| bruising | 0.050 | 0.007 | 0.050 | 0.007 | 0.029 |
| encephalopathy | 0.006 | 0.024 | 0.024 | 0.006 | 0.015 |
| cranberry | 0.057 | 0.006 | 0.057 | 0.006 | 0.032 |
| simvastatin | 0.006 | 0.039 | 0.039 | 0.006 | 0.023 |
| licorice | 0.036 | 0.006 | 0.036 | 0.006 | 0.021 |
| stroke | 0.140 | 0.006 | 0.140 | 0.006 | 0.073 |
| liquorice | 0.005 | 0.012 | 0.012 | 0.005 | 0.008 |
| sertraline | 0.007 | 0.005 | 0.007 | 0.005 | 0.006 |
| diarrhea | 0.160 | 0.005 | 0.160 | 0.005 | 0.082 |
| hydroxyzine | 0.004 | 0.009 | 0.009 | 0.004 | 0.007 |
| hallucinations | 0.030 | 0.004 | 0.030 | 0.004 | 0.017 |
| seizures | 0.108 | 0.004 | 0.108 | 0.004 | 0.056 |
| migraine | 0.043 | 0.004 | 0.043 | 0.004 | 0.023 |
| lithium | 0.010 | 0.003 | 0.010 | 0.003 | 0.007 |
| sweating | 0.033 | 0.003 | 0.033 | 0.003 | 0.018 |

Transitive Closure (proximity) Networks Results

| | psoriasis | insulin glargine | (max) | (min) | (avg) |
|---|---|---|---|---|---|
| allopurinol | 0.072 | 0.053 | 0.072 | 0.053 | 0.063 |
| anaphylaxis | 0.048 | 0.048 | 0.048 | 0.048 | 0.048 |
| primidone | 0.061 | 0.048 | 0.061 | 0.048 | 0.054 |
| sevelamer | 0.061 | 0.048 | 0.061 | 0.048 | 0.054 |
| desmopressin | 0.061 | 0.048 | 0.061 | 0.048 | 0.054 |
| rhinitis | 0.048 | 0.056 | 0.056 | 0.048 | 0.052 |
| bethanechol | 0.062 | 0.047 | 0.062 | 0.047 | 0.055 |
| cabergoline | 0.062 | 0.047 | 0.062 | 0.047 | 0.055 |
| hypersensitivity | 0.044 | 0.059 | 0.059 | 0.044 | 0.052 |

| | | | | | |
|---|---|---|---|---|---|
| pharyngitis | 0.044 | | 0.046 | 0.046 | 0.044 | 0.045 |

---

Query 1W:`**psoriasis**`,`**stroke**`
Ordering: **min**

Proximity Network Results

|  | psoriasis | stroke | (max) | (min) | (avg) |
|---|---|---|---|---|---|
| infections | 0.183 | 0.213 | 0.213 | 0.183 | 0.198 |
| diarrhea | 0.160 | 0.155 | 0.160 | 0.155 | 0.158 |
| constipation | 0.145 | 0.141 | 0.145 | 0.141 | 0.143 |
| psoriasis | 1.000 | 0.140 | 1.000 | 0.140 | 0.570 |
| stroke | 0.140 | 1.000 | 1.000 | 0.140 | 0.570 |
| hepatitis | 0.144 | 0.122 | 0.144 | 0.122 | 0.133 |
| glaucoma | 0.135 | 0.109 | 0.135 | 0.109 | 0.122 |
| seizures | 0.108 | 0.189 | 0.189 | 0.108 | 0.148 |
| hypertension | 0.159 | 0.107 | 0.159 | 0.107 | 0.133 |
| acne | 0.119 | 0.099 | 0.119 | 0.099 | 0.109 |
| infertility | 0.094 | 0.121 | 0.121 | 0.094 | 0.108 |
| vomiting | 0.091 | 0.131 | 0.131 | 0.091 | 0.111 |
| osteoporosis | 0.102 | 0.090 | 0.102 | 0.090 | 0.096 |
| allergic | 0.088 | 0.100 | 0.100 | 0.088 | 0.094 |
| weight loss | 0.085 | 0.088 | 0.088 | 0.085 | 0.087 |
| itching | 0.109 | 0.080 | 0.109 | 0.080 | 0.094 |
| appetite | 0.076 | 0.111 | 0.111 | 0.076 | 0.093 |
| nausea | 0.076 | 0.123 | 0.123 | 0.076 | 0.099 |
| ibuprofen | 0.076 | 0.074 | 0.076 | 0.074 | 0.075 |
| rash | 0.071 | 0.078 | 0.078 | 0.071 | 0.075 |

Transitive Closure (proximity) Networks Results

|  | psoriasis | stroke | (max) | (min) | (avg) |
|---|---|---|---|---|---|
| infections | 0.183 | 0.213 | 0.213 | 0.183 | 0.198 |
| diarrhea | 0.160 | 0.155 | 0.160 | 0.155 | 0.158 |
| constipation | 0.145 | 0.141 | 0.145 | 0.141 | 0.143 |
| psoriasis | 1.000 | 0.140 | 1.000 | 0.140 | 0.570 |
| stroke | 0.140 | 1.000 | 1.000 | 0.140 | 0.570 |
| hepatitis | 0.144 | 0.122 | 0.144 | 0.122 | 0.133 |
| acne | 0.119 | 0.115 | 0.119 | 0.115 | 0.117 |
| glaucoma | 0.135 | 0.109 | 0.135 | 0.109 | 0.122 |
| seizures | 0.108 | 0.189 | 0.189 | 0.108 | 0.148 |
| hypertension | 0.159 | 0.107 | 0.159 | 0.107 | 0.133 |

---

Query 1W:`**glaucoma**`
Ordering: **min**

Proximity Network Results

|  | glaucoma | (max) | (min) | (avg) |
|---|---|---|---|---|

|                       |       |       |       |       |
|-----------------------|-------|-------|-------|-------|
| glaucoma              | 1.000 | 1.000 | 1.000 | 1.000 |
| hepatitis             | 0.195 | 0.195 | 0.195 | 0.195 |
| hypertension          | 0.157 | 0.157 | 0.157 | 0.157 |
| psoriasis             | 0.135 | 0.135 | 0.135 | 0.135 |
| diarrhea              | 0.120 | 0.120 | 0.120 | 0.120 |
| hypotension           | 0.111 | 0.111 | 0.111 | 0.111 |
| stroke                | 0.109 | 0.109 | 0.109 | 0.109 |
| dermatitis            | 0.106 | 0.106 | 0.106 | 0.106 |
| montelukast           | 0.104 | 0.104 | 0.104 | 0.104 |
| vomiting              | 0.101 | 0.101 | 0.101 | 0.101 |
| seizures              | 0.101 | 0.101 | 0.101 | 0.101 |
| acetylsalicylic acid  | 0.095 | 0.095 | 0.095 | 0.095 |
| drowsiness            | 0.090 | 0.090 | 0.090 | 0.090 |
| allopurinol           | 0.079 | 0.079 | 0.079 | 0.079 |
| infections            | 0.078 | 0.078 | 0.078 | 0.078 |
| itching               | 0.077 | 0.077 | 0.077 | 0.077 |
| irritability          | 0.076 | 0.076 | 0.076 | 0.076 |
| constipation          | 0.074 | 0.074 | 0.074 | 0.074 |
| osteoporosis          | 0.074 | 0.074 | 0.074 | 0.074 |
| lidocaine             | 0.073 | 0.073 | 0.073 | 0.073 |

Transitive Closure (proximity) Networks Results

|              | glaucoma | (max) | (min) | (avg) |
|--------------|----------|-------|-------|-------|
| glaucoma     | 1.000    | 1.000 | 1.000 | 1.000 |
| hepatitis    | 0.195    | 0.195 | 0.195 | 0.195 |
| hypertension | 0.157    | 0.157 | 0.157 | 0.157 |
| psoriasis    | 0.135    | 0.135 | 0.135 | 0.135 |
| diarrhea     | 0.120    | 0.120 | 0.120 | 0.120 |
| hypotension  | 0.111    | 0.111 | 0.111 | 0.111 |
| stroke       | 0.109    | 0.109 | 0.109 | 0.109 |
| dermatitis   | 0.106    | 0.106 | 0.106 | 0.106 |
| montelukast  | 0.104    | 0.104 | 0.104 | 0.104 |
| vomiting     | 0.101    | 0.101 | 0.101 | 0.101 |

---

Query 1W:`**glaucoma**`,`**psoriasis**`
Ordering: **min**

Proximity Network Results

|              | psoriasis | glaucoma | (max) | (min) | (avg) |
|--------------|-----------|----------|-------|-------|-------|
| hypertension | 0.159     | 0.157    | 0.159 | 0.157 | 0.158 |
| hepatitis    | 0.144     | 0.195    | 0.195 | 0.144 | 0.170 |
| glaucoma     | 0.135     | 1.000    | 1.000 | 0.135 | 0.567 |
| psoriasis    | 1.000     | 0.135    | 1.000 | 0.135 | 0.567 |
| diarrhea     | 0.160     | 0.120    | 0.160 | 0.120 | 0.140 |
| stroke       | 0.140     | 0.109    | 0.140 | 0.109 | 0.124 |
| dermatitis   | 0.125     | 0.106    | 0.125 | 0.106 | 0.116 |
| seizures     | 0.108     | 0.101    | 0.108 | 0.101 | 0.104 |
| vomiting     | 0.091     | 0.101    | 0.101 | 0.091 | 0.096 |

```
montelukast            0.083    0.104  0.104  0.083  0.093
infections             0.183    0.078  0.183  0.078  0.131
itching                0.109    0.077  0.109  0.077  0.093
irritability           0.112    0.076  0.112  0.076  0.094
constipation           0.145    0.074  0.145  0.074  0.110
osteoporosis           0.102    0.074  0.102  0.074  0.088
acetylsalicylic acid   0.073    0.095  0.095  0.073  0.084
infertility            0.094    0.067  0.094  0.067  0.081
nausea                 0.076    0.067  0.076  0.067  0.071
irritation             0.105    0.064  0.105  0.064  0.084
allergic               0.088    0.062  0.088  0.062  0.075

Transitive Closure (proximity) Networks Results
              psoriasis  glaucoma  (max)  (min)  (avg)
hypertension    0.159     0.157   0.159  0.157  0.158
hepatitis       0.144     0.195   0.195  0.144  0.170
glaucoma        0.135     1.000   1.000  0.135  0.567
psoriasis       1.000     0.135   1.000  0.135  0.567
diarrhea        0.160     0.120   0.160  0.120  0.140
stroke          0.140     0.109   0.140  0.109  0.124
dermatitis      0.125     0.106   0.125  0.106  0.116
seizures        0.108     0.101   0.108  0.101  0.104
vomiting        0.091     0.101   0.101  0.091  0.096
irritability    0.112     0.091   0.112  0.091  0.101
```

---

Query 1W:`glaucoma`,`psoriasis`, `hepatitis`
Ordering: **min**

```
Proximity Network Results
                     psoriasis  glaucoma  hepatitis  (max)  (min)  (avg)
hypertension           0.159     0.157     0.171    0.171  0.157  0.162
hepatitis              0.144     0.195     1.000    1.000  0.144  0.446
glaucoma               0.135     1.000     0.195    1.000  0.135  0.443
psoriasis              1.000     0.135     0.144    1.000  0.135  0.426
diarrhea               0.160     0.120     0.169    0.169  0.120  0.150
stroke                 0.140     0.109     0.122    0.140  0.109  0.124
dermatitis             0.125     0.106     0.155    0.155  0.106  0.129
seizures               0.108     0.101     0.163    0.163  0.101  0.124
vomiting               0.091     0.101     0.126    0.126  0.091  0.106
montelukast            0.083     0.104     0.094    0.104  0.083  0.094
infections             0.183     0.078     0.122    0.183  0.078  0.128
itching                0.109     0.077     0.105    0.109  0.077  0.097
irritability           0.112     0.076     0.145    0.145  0.076  0.111
constipation           0.145     0.074     0.124    0.145  0.074  0.115
osteoporosis           0.102     0.074     0.114    0.114  0.074  0.097
acetylsalicylic acid   0.073     0.095     0.113    0.113  0.073  0.094
infertility            0.094     0.067     0.113    0.113  0.067  0.091
nausea                 0.076     0.067     0.080    0.080  0.067  0.074
```

| | | | | | | |
|---|---|---|---|---|---|---|
| irritation | | 0.105 | 0.064 | 0.100 0.105 0.064 0.090 | | |
| allergic | | 0.088 | 0.062 | 0.084 0.088 0.062 0.078 | | |

Transitive Closure (proximity) Networks Results

| | psoriasis | glaucoma | hepatitis | (max) | (min) | (avg) |
|---|---|---|---|---|---|---|
| hypertension | 0.159 | 0.157 | 0.171 | 0.171 | 0.157 | 0.162 |
| hepatitis | 0.144 | 0.195 | 1.000 | 1.000 | 0.144 | 0.446 |
| glaucoma | 0.135 | 1.000 | 0.195 | 1.000 | 0.135 | 0.443 |
| psoriasis | 1.000 | 0.135 | 0.144 | 1.000 | 0.135 | 0.426 |
| diarrhea | 0.160 | 0.120 | 0.169 | 0.169 | 0.120 | 0.150 |
| stroke | 0.140 | 0.109 | 0.122 | 0.140 | 0.109 | 0.124 |
| dermatitis | 0.125 | 0.106 | 0.155 | 0.155 | 0.106 | 0.129 |
| seizures | 0.108 | 0.101 | 0.163 | 0.163 | 0.101 | 0.124 |
| vomiting | 0.091 | 0.101 | 0.126 | 0.126 | 0.091 | 0.106 |
| irritability | 0.112 | 0.091 | 0.145 | 0.145 | 0.091 | 0.116 |

# Instagram Post Examples

date: Sun, May 25 2014 @ 09:05
"#notmypic .. Say hello to my new friend! Fluoxetina! Side effects by now are a bit of nausea and inquietude.. Better than zoloft! Yesterday night i started to cry while i was with my 2 friends because my ex, bulimia's stress.. I'm sure they thought i'm crazy so i felt like i had to explain my reasons with one of those friends.. Now i'm terrified of his reaction, he is even a friend of my ex.. Don't know what to expect.. It's so hard telling someone about ED and bulimia ▫▫. I'm also thinking about a b/p session today after 2 days clean, maybe it's not the right solution. Idk. #bulimia #bulimic #mia #ed #edfamily #eatingdisorder #prorecovery #bingepurge #purge #binge #fat #prozac #fluoxetine #depression #meds"

date: Wed, May 13 2015 @ 20:05
"I start fluoxetine (D) tomorrow, the doctor switched me from citalopram to this so let's hope it goes better this time #anxietymeds #depressionmeds #citalopram#fluoxetine (D) #anxiety (AE)#depression"

then, same user on the next day:

"ok so I don't know if it's the tablets that are doing this but I feel the lowest I've ever felt and I'm hoping it's not the tablets. Hopefully it's just a bad day, not that there are many good days▫ I hope tomorrow is a better day for everyone, especially if you are feeling the same way I am. #fluoxetine (D) #depression (AE) #anxiety #depressionmeds #anxietymeds"

date: Wed, Dec 24 2014 @ 10:12
"I've gained a significant amount of weight since this summer... It's hard.. It's too much for someone to ask if I'm okay or not. Screw people.. #me #statistic #unhappy #depression (AE) #fluoxetine (D) #christmaseve #yule #hurt #toomanythoughts"

date: Wed, Feb 05 2014 @ 14:02
"i survived another trip to the clinic, saw a specialist, did a test that explained i'm an INFJ (introvert) which is apperently only 1% of the population. Added risperidone and upped ritalin as well as prozac. considering this keeps me "sane" and able to assimilate into the chaos of everyday life i think this counts as my #100happydays today #findhappinessineachday #bipolar #borderlinepersonalitydisorder #INFJ #manicdepression #goinggovernment #prozac #lamotragine #ritalin #risperidone"

date: Tue, Mar 31 2015 @ 18:03
"Anyone has ever tried Prozac !? If so... please share ur feedback
I'd be thankful xo #Depression #anxiety #Bipolar #Prozac #mood_swings



\end{document}